\documentclass[a4paper,onecolumn,unpublished]{quantumarticle}
\pdfoutput=1

\usepackage[utf8]{inputenc}
\usepackage[english]{babel}
\usepackage[T1]{fontenc}
\usepackage{amsmath,amssymb,amsfonts,mathtools}
\usepackage{graphicx}
\usepackage{xcolor}
\usepackage{booktabs}
\usepackage{algorithm}
\usepackage{algpseudocode}
\usepackage[numbers,sort&compress]{natbib}
\usepackage{hyperref}

\begin{document}

\title{Collision-aware Bayesian mitigation of photon loss in linear optical quantum circuits}

\author{Seungwoo Kim}
\email{swfree831@snu.ac.kr}
\affiliation{Department of Electrical and Computer Engineering, Seoul National University, Seoul, Republic of Korea}

\author{Hyunho Cha}
\email{aiden132435@cml.snu.ac.kr}
\affiliation{Department of Electrical and Computer Engineering, Seoul National University, Seoul, Republic of Korea}

\author{Jungwoo Lee}
\thanks{Corresponding author.}
\email{junglee@snu.ac.kr}
\affiliation{Department of Electrical and Computer Engineering, Seoul National University, Seoul, Republic of Korea}

\author{Seunghyun Ahn}
\email{ash9888@snu.ac.kr}
\affiliation{Department of Electrical and Computer Engineering, Seoul National University, Seoul, Republic of Korea}

\author{Donguk Nam}
\email{donguknam@snu.ac.kr}
\affiliation{Department of Electrical and Computer Engineering, Seoul National University, Seoul, Republic of Korea}


\maketitle

\begin{abstract}
Photon loss is the dominant noise source in linear optical quantum circuits. Each lost photon turns an ideal $n$-photon outcome into a pattern with fewer photons, so the detector counts no longer follow the ideal output distribution, and the lossless events that do follow it become exponentially rare as the photon number grows. Discarding the lossy events (postselection) therefore wastes most of the data at realistic loss rates, and the ideal distribution has to be inferred from the lossy counts themselves. Recycling-based mitigation (Mills \& Mezher, \emph{Quantum} 2026; Quandela's \texttt{photon\_recycling}) reweights lossy samples to recover lossless probabilities, but its estimators are biased and restricted to the no-collision sector, where each output mode holds at most one photon. We instead treat the raw counts as Multinomial observations under per-photon binomial loss and infer the ideal output distribution and the loss rate jointly from a Bayesian posterior, which extends naturally to outputs with collisions. On data from the independent Perceval simulator at the scale of prior work ($n=4$ photons in $m=20$ and $m=10$ modes), the posterior has lower KL divergence than both recycling and postselection at every tested loss rate in the no-collision benchmark. In the collision regime it recovers the loss rate and the collision-pattern probabilities, which recycling cannot estimate, and stays below recycling at every loss rate and below postselection except at low loss, where the two are comparable. Its advantage over recycling grows with the sample count, and it supplies credible intervals that the baselines lack.
\end{abstract}

\section{Introduction}
\label{sec:1}

Photonic platforms are among the most promising routes to scalable quantum information processing. Because photons couple only weakly to their environment, they preserve quantum coherence over long distances and are the natural carriers for quantum communication and networking; they can be generated and manipulated at room temperature, clocked at high rates, and integrated on mature silicon-photonic fabrication lines, and they underpin measurement- and fusion-based architectures for fault-tolerant quantum computing~\cite{Knill2001,Kok2007,Bartolucci2023,Alexander2025PsiQuantum}. In the near term, Boson Sampling --- passing $n$ indistinguishable photons through an $m$-mode interferometer described by a Haar-random unitary and recording the output occupation pattern~\cite{AaronsonArkhipov2011} --- is the canonical route to a photonic quantum advantage; early demonstrations used small photonic circuits~\cite{Spring2013,Broome2013}, and reconfigurable devices now realize it on chip~\cite{Maring2024Ascella}.

The very property that makes photons excellent carriers --- their weak environmental coupling --- also means they are lost rather than dephased: a photon absorbed or scattered in a source, waveguide, coupler, or detector simply disappears from the computation. Photon loss is therefore the characteristic and dominant error mechanism of linear optical quantum computing, and it accrues at every stage of a device. Its effect is not merely a lower data rate; loss distorts the measured output distribution, because outcomes in which fewer than $n$ photons survive are produced in place of the ideal $n$-photon patterns. State-of-the-art hardware makes the scale concrete: Boson Sampling demonstrations at 76~\cite{Zhong2020Jiuzhang}, 219~\cite{Madsen2022Borealis}, and 3050~\cite{Liu2025Jiuzhang4} detected photons all operate under double-digit per-mode loss budgets. Loss is moreover a resource for the classical adversary --- tensor-network algorithms that exploit precisely this loss now reproduce experimental output distributions to high fidelity~\cite{Oh2024Bubble} --- so that either loss must be suppressed in hardware~\cite{Maring2024Ascella,Alexander2025PsiQuantum} or its effect must be mitigated classically from the lossy detector counts before any claim about the ideal, lossless distribution can be trusted.

The classical-mitigation route was advanced by Mills and Mezher~\cite{MillsMezher2024}, whose \emph{recycling} framework reconstructs the ideal lossless output distribution directly from lossy photon-counting data in the discrete-variable regime. Its guiding insight is that a sample in which some photons were lost still carries information about the lossless distribution: reweighting such lossy samples across loss orders yields a family of estimators (their Algorithms~1--5) for the ideal probability vector $p$, one of which ships in production as Quandela's Perceval \texttt{photon\_recycling} module. Recycling outperforms na\"ive postselection for moderate-to-high loss at moderate sample counts and has become the reference classical mitigator for this setting. Its reach notwithstanding, two limitations bound what recycling can ultimately deliver. The first is statistical: the authors themselves note that all of their estimators are \emph{biased}, and analyse a component of this bias that does not vanish as the sample count $N \to \infty$. Consistent with that analysis, the recycling error in Sec.~\ref{sec:4} stops decreasing with the sample count: at $n=4$, $m=20$ its KL divergence stays at $0.31$--$0.33$ nats ($\eta=0.3$) and $0.48$--$0.53$ nats ($\eta=0.7$) from $N\approx2\times10^4$ to $2\times10^5$ (Sec.~\ref{sec:4.nsweep}); we do not estimate its asymptotic limit. The second limitation is structural: the recycling kernel $\pi_y^{\mathrm{NC}}(p, \eta) = \eta^k(1-\eta)^{n-k} \sum_{s \in \mathcal{G}_{\mathrm{NC}}(y)} p(s)$ assumes that both lossless inputs and lossy outcomes live in the no-collision sector $\{0,1\}^m$. Real Boson Sampling output rarely respects this: Hong--Ou--Mandel bunching~\cite{HongOuMandel1987} produces collision states with $s_i \geq 2$ whose contribution is substantial at the mode numbers of current experiments: on our circuits collision patterns carry $71\%$ of the lossless output mass at $n=4, m=10$ and still $45\%$ at $m=20$. Recycling cannot accept collision-output samples at all, and discarding them sacrifices both effective sample size and information.

We address both limitations with a Bayesian framework. Treating raw lossy counts directly as Multinomial observations under the per-photon Bernoulli loss model, we infer the joint posterior of the ideal probabilities $p$ and the loss rate $\eta$ via NUTS sampling~\cite{HoffmanGelman2014NUTS}. The framework subsumes the no-collision recycling kernel as a special case and extends naturally to the collision regime through a per-mode binomial loss factor. It applies the Bayesian-mean-estimation philosophy that Blume-Kohout established for quantum state tomography~\cite{BlumeKohout2010BME} and the maximum-likelihood / POVM-Multinomial framework of Hradil and coworkers~\cite{HradilEtAl2004MLE} to recycling-style photon-loss mitigation in linear optical quantum circuits.

\paragraph{Contributions.} This paper makes three contributions:

\begin{enumerate}
  \item \textbf{Reduction of the observed recycling reconstruction error.} Recycling estimators carry reconstruction bias from approximations internal to their reconstruction --- the substitution of the interference term by its expectation, which the original authors note is present in all of their recycling estimators (\cite{MillsMezher2024}, \S4.3), together with the ``primarily heuristic'' decay of the exponential-extrapolation variant. Multinomial-likelihood inference avoids those particular reweighting approximations by fitting the forward model directly. On the $n=4, m=20$ headline at $N=10^5$, generated by an independent simulator, the Bayesian posterior wins KL in $15/15$ and MSE in $14/15$ cells of the loss-rate$\times$indistinguishability grid against both recycling and postselection (paired-bootstrap $p<10^{-4}$), sitting below both baselines at every tested loss rate. An $N$-sweep up to $2\times10^5$ (Sec.~\ref{sec:4.nsweep}) shows the gap widening with $N$, but this finite-sample evidence does not by itself establish asymptotic unbiasedness, which would require a population-limit analysis.
  \item \textbf{Collision-regime support.} A natural one-line modification --- replacing the unit edge weights of the no-collision kernel with multinomial coefficients $\prod_i \binom{s_i}{y_i}$ --- extends the framework to the full collision-aware likelihood. Existing recycling methods (M\&M Algorithms~1--5) cannot operate in this regime at all: forced to discard collision samples and scored on the no-collision subspace, they show substantially higher reconstruction error, while the collision-aware posterior tracks far lower in KL and matches the simulated loss rate to the reported precision across our $(n=4, m=10)$ benchmark (Sec.~\ref{sec:4.3}).
  \item \textbf{Actionable uncertainty quantification.} The joint posterior supplies credible intervals on both the reconstructed distribution $p$ and the loss rate $\eta$ --- error bars that no recycling or postselection baseline provides. In our simulations, where the true distribution is known, retrospective empirical-coverage analysis reveals undercoverage of the ideal distribution at the adverse low-loss, low-indistinguishability corner even though the point estimate remains accurate in KL; there the intervals cover the partially distinguishable distribution that the data determine and miss the ideal one by the distinguishability shift. This is an evaluation diagnostic, not a self-diagnostic available on hardware without ground truth. Point-estimate accuracy (KL) remains our primary metric; the empirical coverage ($0.88$--$0.96$ on the no-collision headline) is reported as a property at the tested resolutions, not a formal simulation-based-calibration guarantee~\cite{Talts2018SBC}.
\end{enumerate}

We restrict experiments to the scaled $n=4$ regime that matches Mills and Mezher's demonstration scale; App.~\ref{app:implementation} develops the complexity analysis and Sec.~\ref{sec:6} discusses approximate-Bayesian routes to larger $n$. The paper proceeds as follows: Sec.~\ref{sec:2} reviews the relevant background in photonic mitigation, recycling, general quantum error mitigation, and Bayesian state estimation; Sec.~\ref{sec:3} develops the joint Bayesian model with the forward kernel and baseline definitions; Sec.~\ref{sec:4} reports head-to-head comparisons at two scaled configurations under independent-simulator noise; Sec.~\ref{sec:5} relates the observed recycling reconstruction-error plateau to known approximation and model-mismatch effects; Sec.~\ref{sec:6} concludes and outlines extensions.

\section{Background and Related Work}
\label{sec:2}

\subsection{Linear optical quantum computing and photon loss}
\label{sec:2.1}

Boson Sampling, introduced by Aaronson and Arkhipov~\cite{AaronsonArkhipov2011}, passes $n$ identical photons through an $m$-mode interferometer described by a Haar-random unitary $U$ and records the output occupation pattern. Sampling the resulting distribution is classically hard under standard complexity-theoretic conjectures, making it the canonical near-term route to photonic quantum advantage. Our work treats this discrete-variable Fock-state regime; we acknowledge the continuous-variable Gaussian Boson Sampling variant~\cite{Hamilton2017GBS}, in which squeezed inputs replace single photons and output probabilities become hafnians, as the experimentally dominant cousin.

Hardware has now entered the loss-limited regime. Jiuzhang demonstrated 76-photon GBS with a $10^{14}$-fold classical speedup~\cite{Zhong2020Jiuzhang}; Borealis added time-multiplexed programmability with up to 219 detected photons~\cite{Madsen2022Borealis}; and the most recent milestone reports 3050-photon GBS with explicit loss-aware defenses against tensor-network spoofing~\cite{Liu2025Jiuzhang4}. The classical side has caught up in step: the tensor-network sampler of Oh et al. shows that simulation complexity \emph{decreases} with loss, so that lossy GBS distributions can be matched or surpassed classically under realistic noise~\cite{Oh2024Bubble}. More generally, loss and partial distinguishability bear on the classical simulability of Boson Sampling: hardness arguments extend to lossy settings~\cite{AaronsonBrod2016}, classical algorithms exploit loss~\cite{OszmaniecBrod2018,GarciaPatron2019} and partial distinguishability~\cite{Renema2018}, and exact and approximate classical samplers~\cite{CliffordClifford2018,Neville2017} set the cost baseline for experiments of the size considered here. Discrete-variable platforms operate in the same loss-budget-bound regime: Quandela's Ascella platform supports up to six-photon Boson Sampling on a reconfigurable interferometer~\cite{Maring2024Ascella}, and PsiQuantum's manufacturable silicon-photonics platform reports high-fidelity qubit preparation, two-photon interference, fusion and chip-to-chip interconnects---conditional on photon detection and not accounting for loss---while previewing low-loss silicon-nitride waveguides and fast electro-optic switches to address loss~\cite{Alexander2025PsiQuantum}. These settings---matching our $n=4$, $m\in\{10,20\}$ experiments---are where classical post-processing of lossy counts most directly affects what can be claimed about the lossless distribution.

\subsection{Recycling-based mitigation and general quantum error mitigation}
\label{sec:2.2}

The most direct precursor to our work is the recycling framework of Mills and Mezher~\cite{MillsMezher2024}. Given lossy $n$-photon counts on $m$ modes, they introduce a family of estimators (their Algorithms 1--5) built around the recycled probability $\tilde{p}^k_R$ (Algorithm~1), which serves as the common building block for the subsequent mitigation routines. Two of these are relevant here: Algorithm~3, which solves a linear system relating recycled probabilities at adjacent photon-loss orders with a dependency parameter $d_k$ that can be either derived from a given $\eta$ or estimated from the empirical deviation ratio (Appendix~12.3.1 of~\cite{MillsMezher2024}); and Algorithm~5, exponential extrapolation across loss orders, which ships in production as Quandela's \texttt{photon\_recycling} module (the implementation we benchmark against). The authors explicitly note that their estimators are biased. They also show that the number of samples needed to construct a recycled probability at loss order $k$ to fixed accuracy scales exponentially with system size, as $O\big(1/[\binom{n}{k}(1-\eta)^{n-k}\eta^{k}]\big)$, and derive conditions on the total sample count up to which recycling outperforms postselection. M\&M's recycling kernel $\Pr(y \mid s, \eta) = \eta^k (1-\eta)^{n-k}$ implicitly restricts the input pattern set $s$ to no-collision strings; we revisit this assumption in Sec.~\ref{sec:5.1}.

Earlier photonic-specific mitigation appeared in the continuous-variable GBS setting, where Su et al. proposed two loss-mitigation schemes built on additional data collection and classical post-processing~\cite{SuEtAl2021PhotonicQEM}. In the broader circuit-model literature, zero-noise extrapolation (ZNE) is the canonical paradigm, introduced in parallel by Temme, Bravyi and Gambetta~\cite{TemmeBravyiGambetta2017} and by Li and Benjamin~\cite{LiBenjamin2017}; Mills and Mezher show that ZNE yields unbiased estimators under photon loss but, because the underlying Vandermonde inversion amplifies statistical error, provide evidence that it cannot improve on postselection at any loss rate in the discrete-variable LOQC setting. Virtual distillation offers a different bias-suppression route via ancilla circuits and noise dilution~\cite{TeoEtAl2023VirtualDistillation}. The unifying review of Cai et al.~\cite{CaiEtAl2023RMPReview} frames quantum error mitigation as a statistical-inference problem, a perspective our Bayesian formulation makes explicit; the earlier review of Endo et al. surveys error extrapolation, probabilistic error cancellation (quasi-probability methods), symmetry verification and quantum subspace expansion in NISQ contexts~\cite{EndoEtAl2021Review}. The closest qubit-setting analogue of recycling is readout-error mitigation, where measured frequencies are corrected by inverting a calibrated noise matrix~\cite{Bravyi2021} or by iterative Bayesian unfolding~\cite{Nachman2020}; unlike these calibrated-noise approaches, we infer the loss rate jointly with the distribution. Takagi et al. establish fundamental sampling-overhead bounds for mitigation protocols, quantifying the bias--variance trade-offs against which likelihood-based Bayesian estimation should be assessed~\cite{TakagiEtAl2022Limits}.

\section{Method}
\label{sec:3}

\subsection{Forward model: lossy probabilities under collision}
\label{sec:3.1}

Consider $n$ indistinguishable single photons injected into an $m$-mode passive linear-optical interferometer realising a Haar-random unitary $U$, the standard Aaronson--Arkhipov setting for Boson Sampling~\cite{AaronsonArkhipov2011}. In the lossless limit the device produces an output state $s \in \mathbb{Z}_{\geq 0}^m$ with $\sum_i s_i = n$ drawn from the ideal distribution $p(s)$, whose values are squared permanents of submatrices of $U$. Throughout, $p$ denotes the target of inference. Photon loss is modelled, following Mills and Mezher~\cite{MillsMezher2024} and the broader photonic-QEM literature~\cite{SuEtAl2021PhotonicQEM,Maring2024Ascella}, as an independent per-photon Bernoulli channel with uniform survival probability $1-\eta$. After loss, $k = n - \sum_i y_i$ photons are missing from the detected outcome $y \in \mathbb{Z}_{\geq 0}^m$ and the surviving photons are still distributed across the modes of $U$.

\paragraph{No-collision (NC) regime.} M\&M's recycling framework restricts both lossless input $s$ and lossy output $y$ to binary vectors. In this regime the input sector has size $K_{\mathrm{NC}} = \binom{m}{n}$ and the output sector is $\mathcal{Y}_{\mathrm{NC}} = \{y \in \{0,1\}^m : \sum_i y_i \leq n\}$. Conditional on a lossless state $s$, the probability of observing a particular NC outcome $y \leq s$ is $\eta^{k}(1-\eta)^{n-k}$: the $k$ lost photons must be exactly the ones in the modes where $s_i = 1$ and $y_i = 0$. Marginalising over $s$ with the ideal $p$ gives
\begin{equation}
\label{eq:nc-likelihood}
\pi_y(p, \eta) \;=\; \eta^{k}\,(1-\eta)^{n-k}
  \sum_{s \in \mathcal{G}(y)} p(s),
\qquad
k = n - \textstyle\sum_i y_i,
\end{equation}
with $\mathcal{G}(y) = \{s \in \{0,1\}^m : \sum_i s_i = n,\ s \geq y\}$. Equation~\eqref{eq:nc-likelihood} coincides with the NC recycling kernel of~\cite{MillsMezher2024}; we use it as a forward model rather than inverting it estimator-by-estimator (Sec.~\ref{sec:3.2}).

\paragraph{Collision-aware extension.} Real Boson Sampling output rarely respects the NC restriction: at the moderate $m/n^2$ values used in current hardware~\cite{MillsMezher2024,Maring2024Ascella}, Hong--Ou--Mandel bunching~\cite{HongOuMandel1987} places a non-trivial share $f_{\mathrm{coll}}$ of the lossless output mass on collision patterns --- $0.71$ at $(n,m)=(4,10)$ and $0.45$ at $(4,20)$, averaged over our $30$ Haar circuits. Loss lowers the fraction of \emph{observed} samples that still show a collision ($0.41$ at $m=10$, $\eta=0.3$), but the lossy outcomes of collision patterns remain in the data. We accommodate collisions by expanding the pattern space to all weak compositions
\[
\mathcal{P}_{\mathrm{coll}} = \bigl\{s \in \mathbb{Z}_{\geq 0}^m : \textstyle\sum_i s_i = n\bigr\}, \qquad K_{\mathrm{coll}} = \binom{m+n-1}{n}.
\]
Because lossless photons within the same mode are exchangeable, the number of photons surviving in mode $i$ given $s_i$ occupants follows a Binomial channel,
\begin{equation}
\label{eq:per-mode-binomial}
\Pr(y_i \mid s_i, \eta) \,=\, \binom{s_i}{y_i}(1-\eta)^{y_i}\eta^{s_i-y_i}, \qquad y_i \leq s_i,
\end{equation}
independent across modes. Marginalising the resulting product over $s \in \mathcal{G}_{\mathrm{coll}}(y) = \{s \in \mathcal{P}_{\mathrm{coll}} : s \geq y\}$ yields
\begin{equation}
\label{eq:coll-likelihood}
\pi_y(p, \eta) \;=\; \eta^{k}\,(1-\eta)^{n-k}
  \sum_{s \in \mathcal{G}_{\mathrm{coll}}(y)}
  \Biggl[\prod_{i=1}^{m} \binom{s_i}{y_i}\Biggr] p(s).
\end{equation}
When $s, y \in \{0,1\}^m$ every $\binom{s_i}{y_i} \in \{1\}$, so~\eqref{eq:coll-likelihood} reduces to~\eqref{eq:nc-likelihood}: the NC kernel is the boundary case of the collision-aware likelihood rather than a separate model. The product of binomial coefficients $\prod_i \binom{s_i}{y_i}$ is the single structural element that the recycling family of~\cite{MillsMezher2024} omits; restoring it corrects this collision-regime model mismatch (Sec.~\ref{sec:5.1}), which is distinct from the reconstruction approximations used by recycling in the no-collision regime. Inference for both regimes uses the same Bayesian machinery developed in Sec.~\ref{sec:3.2}, with implementation details (sparse incidence representation, NUTS sampler) deferred to App.~\ref{app:implementation}.

\subsection{Bayesian framework}
\label{sec:3.2}

Given $N$ independent samples summarised by counts $\mathbf{c} = (c(y))_{y \in \mathcal{Y}}$, the data likelihood is Multinomial in $\pi(p, \eta)$. We use symmetric reference priors:
\begin{align}
p &\,\sim\, \mathrm{Dirichlet}(\alpha \mathbf{1}_K), \quad \alpha = 1, \label{eq:prior-p}\\
\eta &\,\sim\, \mathrm{Beta}(\beta, \beta), \quad \beta = 1, \label{eq:prior-eta}\\
\bigl(c(y_1),\dots,c(y_{|\mathcal{Y}|})\bigr) \mid p, \eta &\,\sim\, \mathrm{Multinomial}\!\bigl(N,\ \pi(p,\eta)\bigr), \label{eq:lik-multinomial}
\end{align}
where the same $\pi_y(p, \eta)$ is the NC kernel~\eqref{eq:nc-likelihood} in the no-collision regime or the collision-aware kernel~\eqref{eq:coll-likelihood} otherwise. These priors are modelling choices, not conjugate priors for this transformed likelihood: $p$ enters through the loss map $\pi=A(\eta)p$, while $\eta$ also changes the mixture weights. Moreover, although $\alpha=1$ is uniform with respect to Lebesgue measure on the simplex, it need not be weakly informative for every functional when $K$ is large; App.~\ref{app:prior} therefore reports a sensitivity analysis. Bayes' theorem yields the joint posterior
\begin{equation}
\label{eq:posterior}
\Pr(p, \eta \mid \mathbf{c}) \;\propto\; \prod_{y \in \mathcal{Y}} \pi_y(p, \eta)^{c(y)}\ \cdot\ \Pr(p)\ \cdot\ \Pr(\eta).
\end{equation}

\paragraph{From per-pattern recycling to a joint likelihood.} The distinction between this paper and~\cite{MillsMezher2024} is the role of the forward kernel $\pi_y$ within the inference procedure. M\&M treat $\pi_y$ as a per-pattern frequentist estimator, deriving Hoeffding-type concentration bounds on the recycled probability $\tilde{p}^k_R(s)$ at each loss order $k$ and combining those bounds across $k$ via linear solving (their Algorithm~3) or exponential extrapolation (their Algorithm~5, shipped as Quandela \texttt{photon\_recycling}). Inversely, we treat $\pi_y$ as the data-generating likelihood of the full Multinomial count vector $\mathbf{c}$ and apply Bayes' theorem once over the joint $(p, \eta)$ space. For i.i.d.\ shots the count vector $\mathbf{c}$ is a sufficient statistic, and the Multinomial likelihood uses all of it jointly through the shared parameters $(p, \eta)$. M\&M's estimators also act on $\mathbf{c}$, but they estimate each loss order separately and combine the results by approximate reweighting rather than through a joint likelihood, so they do not use the coupling of the loss orders through common $(p, \eta)$; postselection, and any no-collision restriction (including our own no-collision track, App.~\ref{app:coll-m20}), additionally discard samples outright. The recycling estimators exhibit the reconstruction-error plateau observed in Sec.~\ref{sec:4} and analysed in relation to their reported bias in Sec.~\ref{sec:5.1}. The collision-aware kernel~\eqref{eq:coll-likelihood} further generalises a no-collision restriction that the recycling family implicitly carries. The vocabulary of the present framework --- likelihood, Multinomial, Bayesian, posterior, prior, Dirichlet, credible interval --- does not appear in the M\&M text, which operates entirely within a frequentist concentration-bound framework.

\paragraph{Consequences.} Three properties of the inference procedure follow from this shift. First, $p$ and $\eta$ are inferred jointly from a single posterior, and a marginal posterior for $\eta$ is available as a byproduct; recycling baselines either ignore $\eta$ (postselection), absorb it implicitly during estimator construction (Quandela \texttt{photon\_recycling}; the data-driven variant of M\&M Algorithm~3), or require it as an exogenous input (M\&M Algorithm~3 in its $\eta$-given form), with no posterior on $\eta$ in any case. Second, working directly with the Multinomial likelihood follows the POVM/Multinomial framework formalised for quantum measurement by Hradil and collaborators~\cite{HradilEtAl2004MLE}, and inherits the uncertainty-quantification advantages identified by Blume-Kohout for Bayesian mean estimation over maximum-likelihood point estimates~\cite{BlumeKohout2010BME}. Third, the posterior is sampled via the No-U-Turn Sampler (NUTS)~\cite{HoffmanGelman2014NUTS} on a sparse incidence representation of the kernel that kept inference tractable up to $K=8855$ at $N=10^5$ in our runs; the standard tools and their composition for this likelihood are documented in App.~\ref{app:implementation}. Together these properties support the finite-sample error reduction and empirical credible-interval coverage reported in Sec.~\ref{sec:4}; they do not alone imply asymptotic unbiasedness.

\subsection{Baseline methods}
\label{sec:3.3}

Two baselines anchor the empirical comparison. \emph{Postselection} discards every sample with $\sum_i y_i < n$ (and, in the collision regime, additionally drops samples with $\max_i y_i \geq 2$ via a collision filter applied for apples-to-apples comparison with the NC-restricted competitors) and estimates $p(s)$ by the empirical frequency of the surviving NC outcomes; it is unbiased but its effective sample size shrinks accordingly. \emph{Quandela \texttt{photon\_recycling}} is the official Perceval implementation of M\&M Algorithm~5 (exponential extrapolation across loss orders): lossy outcomes at every $k$ are recycled into recovered probabilities and combined to estimate the lossless distribution, with the estimator restricted to NC inputs. We do not report Mills and Mezher's linear-solving variant (their Algorithm~3), which shares the NC kernel and the interference approximation with Algorithm~5. Both baselines and our Bayesian estimator are evaluated on identical sample paths.

\section{Results}
\label{sec:4}

We evaluate the Bayesian Multinomial estimator at the two scaled configurations that match the experimental scale of Mills and Mezher~\cite{MillsMezher2024}: the no-collision headline $(n,m)=(4,20)$ with $K_{\mathrm{NC}}=\binom{20}{4}=4845$, and the collision-aware regime $(n,m)=(4,10)$ with $K_{\mathrm{coll}}=\binom{13}{4}=715$. All data are generated by the Strong Linear Optical Simulation (SLOS) engine~\cite{Heurtel2023SLOS} of Perceval~\cite{Heurtel2023Perceval} under its physical \texttt{NoiseModel} --- transmittance $1-\eta$, photon indistinguishability $V$, and multiphoton $g^{(2)}$ --- while the inference likelihood of Sec.~\ref{sec:3} is left unchanged. This separation has two interpretations, which differ between the two tracks. On the collision-aware track, at $V=1$ and $g^{(2)}=0$ generator and likelihood encode the same statistical model, so the experiment is a matched-model cross-check implemented through an independent simulation backend. The no-collision track is not matched even there: bosonic bunching places nearly half of the lossless output (about $45\%$) on collision patterns at $n=4$, $m=20$, which the no-collision likelihood, like the recycling estimators, does not model (App.~\ref{app:coll-m20}). The two remaining noise axes act differently. Multiphoton emission ($g^{(2)}>0$) produces events the single-$\eta$ likelihood cannot represent, a genuine misspecification that also shifts $\hat\eta$ (App.~\ref{app:g2}). Partial distinguishability ($V<1$) does not: loss acts on each photon independently of its internal state, so the data are uniform loss applied to the partially distinguishable lossless distribution $p_V$, which the unstructured $p$ of the likelihood represents exactly. Under $V<1$ the posterior therefore targets $p_V$ rather than the ideal $p^*$, and any estimator that does not model $V$ inherits the gap between them (App.~\ref{app:target-shift}). Every cell reports $R=30$ Haar-random unitary replications at $N=10^5$ samples, paired by seed across methods; primary metrics are KL divergence on the NC subspace (clip--floor--renormalize protocol, Eq.~\eqref{eq:kl-smooth}) and empirical coverage of the $95\%$ posterior credible intervals on $p$.

\begin{equation}
\label{eq:kl-smooth}
\mathrm{KL}\bigl(p^* \,\Vert\, \hat p\bigr)
\,=\, \sum_{s} p^*(s)\,\log\!\frac{p^*(s)}{\tilde p(s)},
\qquad
\tilde p(s) \,=\, \frac{\max\!\bigl(\hat p(s),\ 1/N\bigr)}{\sum_{s'} \max\!\bigl(\hat p(s'),\ 1/N\bigr)}.
\end{equation}

\subsection{Setup: independent-tool generation and the noise axes}
\label{sec:4.1}

Data generation and inference are fully decoupled. For a fixed Haar unitary $U$, Perceval evolves the $n$-photon input through $U$ with SLOS (an exact Fock-amplitude engine, numerically independent of our permanent-plus-closed-form-loss kernel) and samples $N$ detected patterns under the physical noise model; collision and multiphoton events that Perceval produces are outside the NC likelihood's support and are handled exactly as an experimenter would (NC-subspace scoring; $\sum_i y_i>n$ events discarded). Under the NC likelihood a collision event is discarded if a mode still holds two or more photons after loss, and is otherwise attributed to no-collision inputs; App.~\ref{app:coll-m20} quantifies the resulting loss-rate bias. The inference target $p^*$ is the ideal ($V=1$, lossless) distribution on the \emph{same} $U$, so scoring is apples-to-apples.

The indistinguishability $V$ is the primary noise axis, and its interpretation must be stated precisely. At $V=1, g^{(2)}=0$ the Perceval generator reproduces the distribution assumed by the collision-aware likelihood, so on that track this slice is an independent-implementation \emph{cross-check}, not an escape from the inverse crime. For the no-collision likelihood the same slice already carries one mismatch, the omitted collision patterns, which it shares with the recycling baselines; we therefore read the $V=1$ no-collision comparison as a comparison of estimators under a common, known misspecification rather than as a matched test. Further misspecification arises at $g^{(2)}>0$ (multiphoton emission the single-$\eta$ Binomial channel cannot represent). Partial distinguishability $V<1$ instead moves the inference target. It interpolates the statistics away from the bosonic permanent, but the likelihood places no permanent structure on $p$ and represents the $V<1$ data exactly with $p=p_V$. The loss rate therefore remains identified (exactly, on the collision-aware track, at every $V$; Table~\ref{tab:coll-headline}), while the error measured against $p^*$ includes the distance $\mathrm{KL}(p^*\Vert p_V)$, which no estimator that ignores $V$ can remove (App.~\ref{app:target-shift}). The robustness claims below therefore rest on the omitted collisions of the no-collision track and on $g^{(2)}>0$; the $V<1$ results measure how close each estimator stays to $p^*$ when the target has moved, with $V=1$ reported as the baseline for each track.

\subsection{No-collision headline at $n=4, m=20$}
\label{sec:4.2}

\begin{figure}[t]
  \centering
  \includegraphics[width=0.85\textwidth]{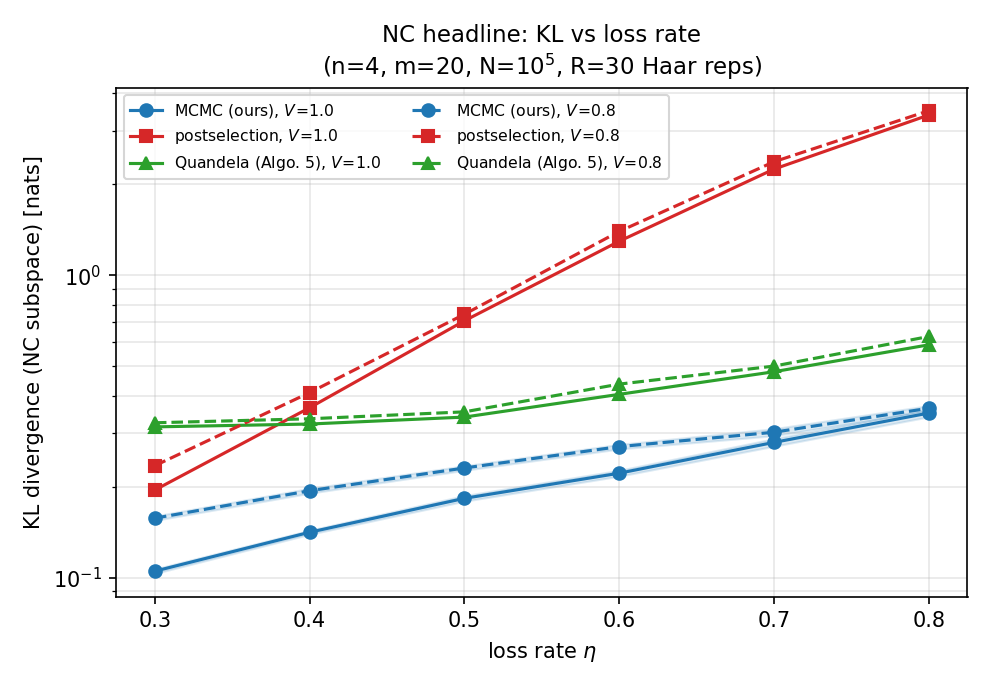}
  \caption{No-collision headline ($n=4, m=20$, $N=10^5$, $R=30$): KL divergence on the NC subspace vs.\ loss rate $\eta$, under Perceval-generated data. Solid = $V=1.0$; dashed = $V=0.8$. Both slices are misspecified for the no-collision likelihood, which omits collision patterns (App.~\ref{app:coll-m20}); at $V=0.8$ the lossless target also moves from $p^*$ to $p_V$ (App.~\ref{app:target-shift}). The Bayesian posterior (blue) is lowest at every $\eta$ under both slices; postselection (red) crosses up as its effective sample count collapses (at high loss its KL mostly reflects the $1/N$ floor applied to its many zero estimates; Sec.~\ref{sec:4.nsweep}), and the recycling estimator (green) exhibits an error plateau at the tested sample budget. Shaded band: $95\%$ percentile-bootstrap CI on MCMC.}
  \label{fig:nc-kl-eta}
\end{figure}

At the headline scale the Bayesian posterior sits below every recycling and postselection baseline at every loss rate, and it does so at both $V=1.0$, where the no-collision model already omits collision patterns, and $V=0.8$, where the lossless target itself moves from $p^*$ to $p_V$ (Fig.~\ref{fig:nc-kl-eta}; App.~\ref{app:target-shift}). Postselection crosses above the Bayesian curve as $N_{\mathrm{eff}}=N(1-\eta)^n$ collapses with loss---at $\eta\geq0.7$ most of its $K$ estimates are exactly zero, so its KL there is dominated by the $1/N$ floor of the KL protocol (Sec.~\ref{sec:4.nsweep}); the recycling estimators (M\&M Algorithm~5, shipped as Quandela \texttt{photon\_recycling}) exhibit a nearly flat reconstruction-error plateau near $0.3$--$0.6$ nats in this fixed-$N$ comparison (Sec.~\ref{sec:5.1}). Table~\ref{tab:nc-headline} lists the metrics: the MCMC KL is lowest in all $6/6$ $(\eta,V)$ cells shown, and across the full $\eta\in\{0.3,0.7,0.9\}\times V\in\{1.0,\dots,0.8\}$ grid MCMC wins KL $15/15$ and MSE $14/15$ against both baselines (paired-bootstrap $p<10^{-4}$). The jointly-inferred loss-rate estimate is $\hat\eta=0.342$ at $\eta=0.3$, $0.716$ at $\eta=0.7$ and $0.902$ at $\eta=0.9$ ($V=1.0$); this bias comes from the collision events that the NC model discards or misattributes, and the collision-aware fit removes it (App.~\ref{app:coll-m20}).

\begin{table}[t]
\centering
\caption{No-collision headline metrics at $n=4, m=20, N=10^5$, KL on the NC subspace. Mean over $R=30$ Haar replications; $95\%$ percentile-bootstrap CI in brackets (paired by seed). \textbf{Bold} = best in row. At $V=1.0$ the only mismatch is the collision patterns omitted by the no-collision likelihood and by the baselines; at $V=0.8$ the lossless target also moves from $p^*$ to $p_V$ (App.~\ref{app:target-shift}). Coverage is of the $95\%$ credible intervals on $p$. At $\eta\geq0.7$ the postselection KL is dominated by the $1/N$ floor applied to its zero estimates (Sec.~\ref{sec:4.nsweep}) and should not be read as an accuracy ratio.}
\label{tab:nc-headline}
\small
\begin{tabular}{ccccccc}
\toprule
$\eta$ & $V$ & \textbf{MCMC KL} & post KL & Algo.\ 5 KL & $\hat\eta$ & cov \\
\midrule
0.3 & 1.0 & $\mathbf{0.105\,[0.103,0.107]}$ & $0.195$ & $0.315$ & $0.342$ & $0.95$ \\
0.3 & 0.8 & $\mathbf{0.157\,[0.155,0.160]}$ & $0.235$ & $0.325$ & $0.338$ & $0.88$ \\
0.7 & 1.0 & $\mathbf{0.280\,[0.272,0.288]}$ & $2.248$ & $0.480$ & $0.716$ & $0.96$ \\
0.7 & 0.8 & $\mathbf{0.303\,[0.295,0.311]}$ & $2.377$ & $0.501$ & $0.714$ & $0.96$ \\
0.9 & 1.0 & $\mathbf{0.413\,[0.405,0.422]}$ & $0.680$ & $0.588$ & $0.902$ & $0.95$ \\
0.9 & 0.8 & $\mathbf{0.417\,[0.408,0.427]}$ & $0.680$ & $0.624$ & $0.902$ & $0.95$ \\
\bottomrule
\end{tabular}
\end{table}

\begin{figure}[t]
  \centering
  \includegraphics[width=0.78\textwidth]{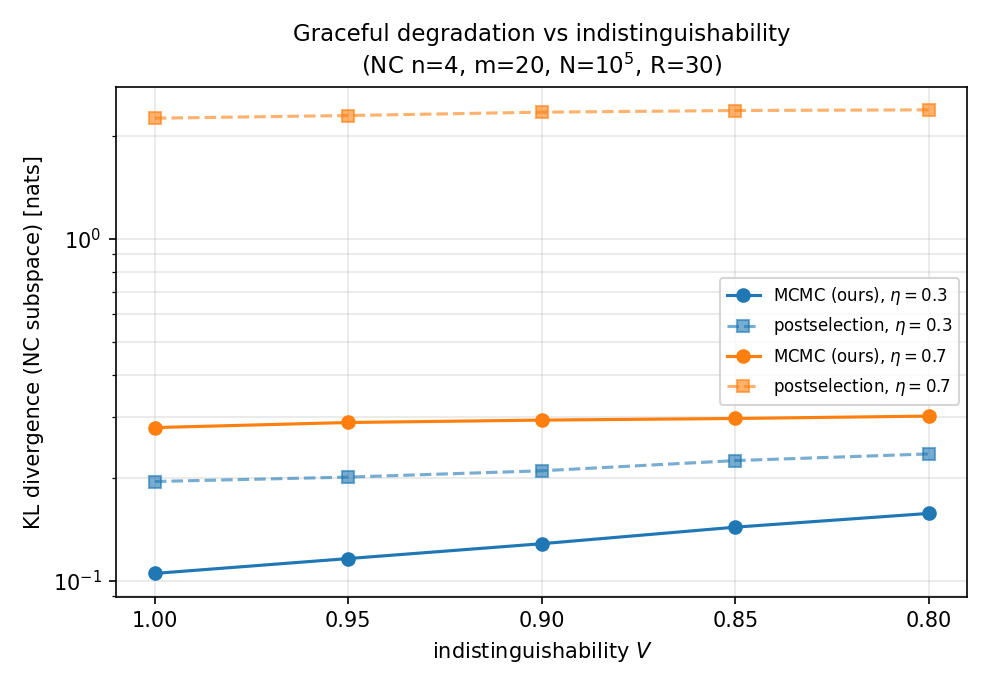}
  \caption{KL against the ideal $p^*$ vs.\ indistinguishability $V$ (NC $n=4, m=20$, $N=10^5$, $R=30$). Both the Bayesian posterior and postselection degrade smoothly as $V$ decreases from $1.0$ to $0.8$; the Bayesian curve keeps its margin. Neither method models $V$, so both target the partially distinguishable distribution $p_V$, and the rise largely reflects the distance between $p_V$ and $p^*$ (App.~\ref{app:target-shift}).}
  \label{fig:nc-kl-V}
\end{figure}

Neither method models $V$, so the $V<1$ comparison measures how close each estimator stays to $p^*$ when the target moves, not distinguishability correction. Figure~\ref{fig:nc-kl-V} makes this explicit: as $V$ falls from $1.0$ to $0.8$ both the Bayesian posterior and postselection degrade smoothly, because the lossless distribution itself moves away from the ideal, by $\mathrm{KL}(p^*\Vert p_V)=0.044$ nats at $V=0.8$ averaged over the circuits (App.~\ref{app:target-shift}), while the Bayesian curve retains its margin throughout. At low loss ($\eta=0.3$) postselection has about twice the posterior's KL ($0.195$ vs.\ $0.105$ at $V=1.0$); at $\eta\geq0.7$ its KL is dominated by the $1/N$ floor applied to its zero estimates (Sec.~\ref{sec:4.nsweep}). The correct reading is therefore that the Bayesian estimator follows the moved target and stays closest to $p^*$, not that it corrects for distinguishability.

\subsection{Dependence on the sample budget}
\label{sec:4.nsweep}

\begin{figure}[t]
  \centering
  \includegraphics[width=0.95\textwidth]{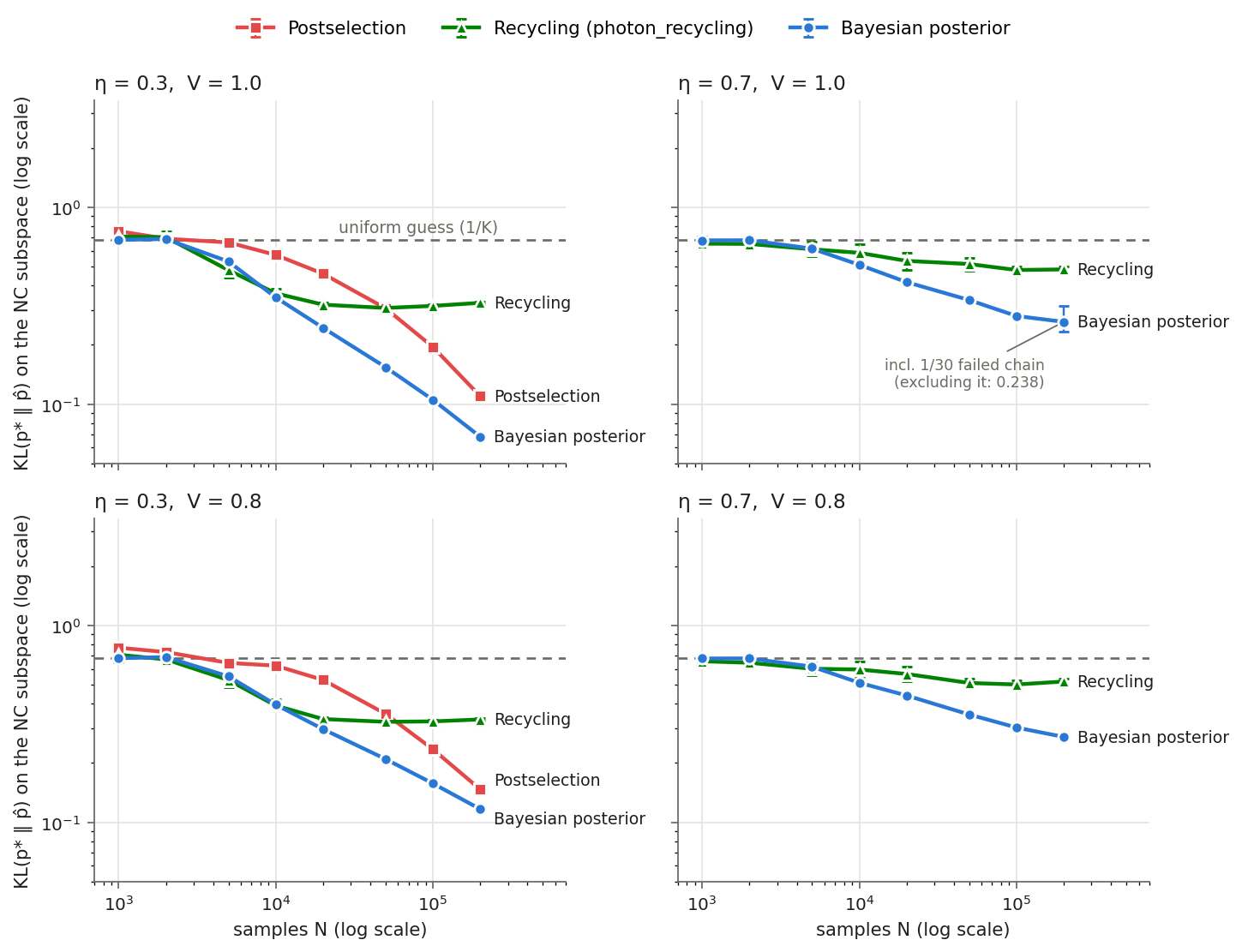}
  \caption{KL divergence vs.\ sample count $N$ on the no-collision track ($n=4, m=20$, Perceval data, the same $R=30$ Haar unitaries at every $N$; the $N=10^5$ points are the headline runs). Points: mean; bars: $95\%$ percentile-bootstrap CI. Dashed: uninformed uniform guess $p=1/K$. Recycling plateaus from $N\approx2\times10^4$ while the posterior keeps improving. Postselection is omitted at $\eta=0.7$, where most of its estimates are zero and its KL is set by the $1/N$ floor of the KL protocol. The $N=2\times10^5$ mean at $\eta=0.7$, $V=1.0$ includes one circuit on which the sampler failed ($0.238$ without it).}
  \label{fig:nc-kl-N}
\end{figure}

To see how the comparison depends on $N$, we repeated the no-collision benchmark at $N\in\{1,2,5\}\times10^3$, $\{1,2,5\}\times10^4$ and $2\times10^5$ on the same $30$ Haar unitaries as the headline runs at $N=10^5$ (Fig.~\ref{fig:nc-kl-N}). At $N\lesssim2\times10^3$ all three estimators are close to an uninformed uniform guess (mean KL $0.680$ over these circuits): the data do not determine $K=4845$ probabilities. Recycling improves until $N\approx2\times10^4$ and then stays essentially flat (KL $0.31$--$0.33$ at $\eta=0.3$ and $0.48$--$0.53$ at $\eta=0.7$, $V=1.0$), which is the plateau discussed in Sec.~\ref{sec:5.1}. The posterior keeps improving (at $\eta=0.3$, $V=1.0$: $0.348$ at $N=10^4$, $0.105$ at $10^5$ and $0.068$ at $2\times10^5$), so the gap to recycling widens with $N$. At $N=10^4$ the two are comparable (the posterior is lower on $17$ of $30$ circuits at $\eta=0.3$, $V=1.0$); the clear advantage of Sec.~\ref{sec:4.2} is a large-$N$ property. Postselection, which is unbiased for the no-collision target, also keeps improving at $\eta=0.3$ ($0.110$ at $2\times10^5$). At $\eta=0.7$ it keeps about $0.4\%$ of the samples, most of its estimates are exactly zero, and its KL rises with $N$ because the $1/N$ floor of the KL protocol penalises those zeros more heavily as $N$ grows; it is therefore not shown there.

\begin{figure}[t]
  \centering
  \includegraphics[width=0.95\textwidth]{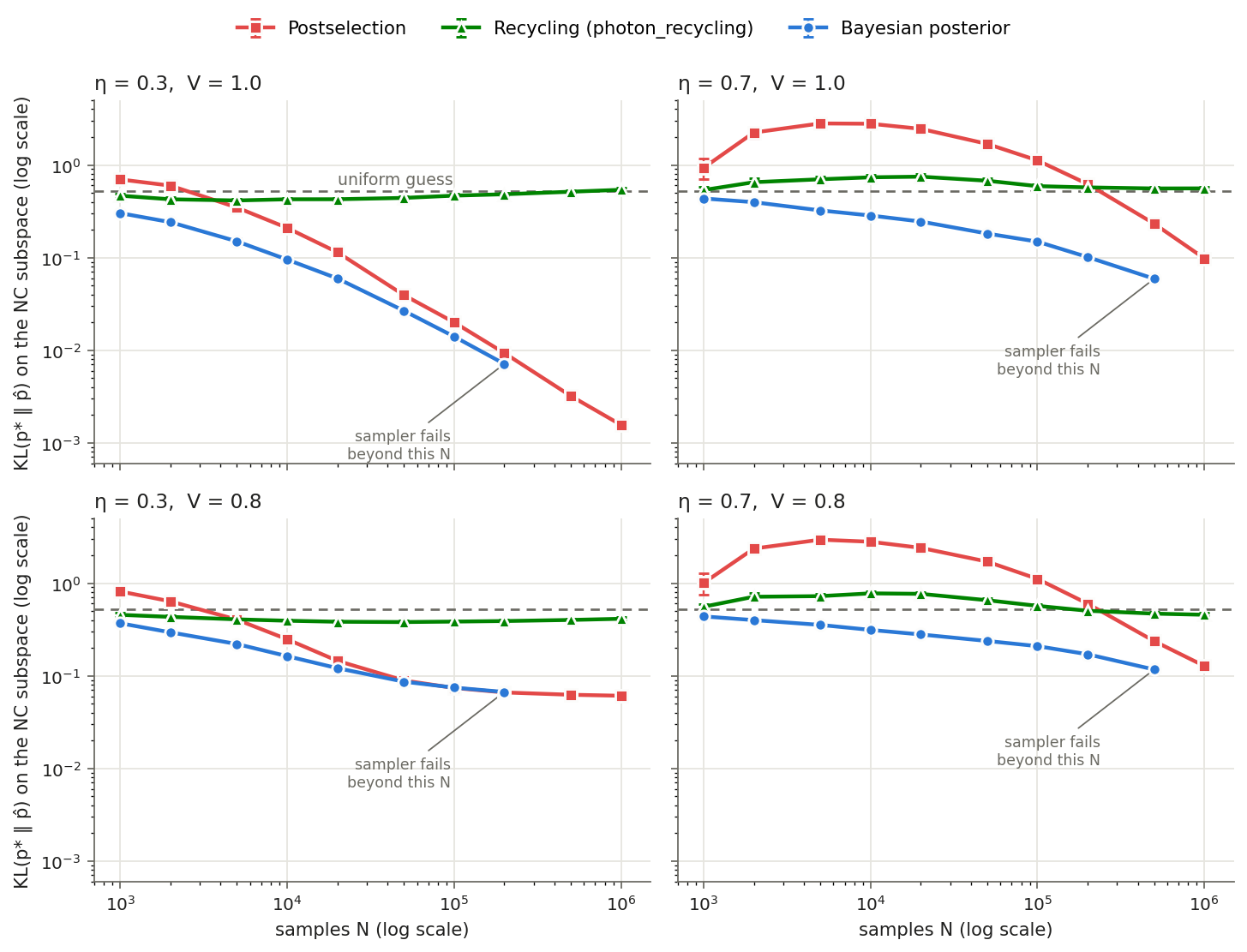}
  \caption{KL divergence vs.\ sample count $N$ on the collision-aware track ($n=4, m=10$, Perceval data, the same $R=30$ Haar unitaries at every $N$; the $N=10^5$ points are the runs of Table~\ref{tab:coll-headline}), on the NC subspace. Points: mean; bars: $95\%$ percentile-bootstrap CI. Dashed: uninformed uniform guess on the NC subspace. The posterior is shown only up to the largest $N$ at which NUTS converged ($2\times10^5$ at $\eta=0.3$, $5\times10^5$ at $\eta=0.7$); the baselines up to $10^6$. At $\eta=0.7$ the postselection KL at small $N$ is set by the $1/N$ floor applied to its zero estimates.}
  \label{fig:coll-kl-N}
\end{figure}

On the collision-aware track we repeated the comparison at $N$ from $10^3$ to $10^6$ on the same $30$ circuits as Table~\ref{tab:coll-headline} (Fig.~\ref{fig:coll-kl-N}). Recycling again stops improving: its KL stays between $0.38$ and $0.78$ nats at every $N$ up to $10^6$. The posterior keeps improving with $N$ (at $\eta=0.7$, $V=1.0$: $0.285$ at $N=10^4$, $0.149$ at $10^5$ and $0.059$ at $5\times10^5$) and is the lowest of the three at $\eta=0.7$ throughout. At $\eta=0.3$ postselection, which is unbiased and keeps about $7\%$ of the samples there, closes the gap as $N$ grows: at $V=1.0$ the posterior is lower up to $N=2\times10^5$ ($0.007$ against $0.009$), and at $V=0.8$ the two coincide from $N\approx5\times10^4$, where both approach the distance between $p_V$ and $p^*$ (App.~\ref{app:target-shift}). Exact NUTS converges over a wider range of $N$ here than at $K=4845$, but not up to $10^6$. Four-chain diagnostics (two circuits per loss rate, production and reference budgets) show convergence at $N=5\times10^5$ for $\eta=0.7$ ($\hat R\leq1.003$ for $\eta$) but not for $\eta=0.3$ ($\hat R$ up to $1.29$), and failure for both at $N=10^6$ ($\hat R$ between $1.7$ and $3.7$, also with the reference budget); in the $30$-circuit runs the coverage in these cells is $0.2$ or below. Figure~\ref{fig:coll-kl-N} therefore shows the posterior only up to $N=2\times10^5$ at $\eta=0.3$ and $5\times10^5$ at $\eta=0.7$.

Two limitations appear in the same sweep. First, the posterior-mean loss rate approaches $0.343$ at $\eta=0.3$ and $0.716$ at $\eta=0.7$ rather than the true values, the signature of the collision mismatch of the no-collision likelihood (Sec.~\ref{sec:5.1}); fitting the collision-aware likelihood to the same circuits removes this bias ($\hat\eta=0.298$ and $0.697$) without lowering the KL (App.~\ref{app:coll-m20}). At $\eta=0.3$ the empirical coverage also falls as $N$ grows: slightly at $V=1.0$ ($0.94$ at $N=2\times10^5$), and markedly at $V=0.8$ ($0.83$), consistent with intervals that narrow below the distance between $p_V$ and $p^*$ (App.~\ref{app:target-shift}). Second, exact NUTS with the budget of App.~\ref{app:implementation.sparse} begins to fail beyond $N=10^5$: at $N=2\times10^5$ the chain fails on one of the $30$ circuits at $\eta=0.7$, $V=1.0$ (coverage $0.09$), and a four-chain check at $N=10^6$ failed outright (every draw at the maximum tree depth, $\hat R=3.0$), so we do not report $N=10^6$; the collision-aware track shows the same failure at $N=10^6$ (see above). A sampler that scales to larger $N$, such as data-augmentation Gibbs sampling over the latent loss assignments, is left for future work.

\subsection{Collision-aware regime at $n=4, m=10$}
\label{sec:4.3}

\begin{figure}[t]
  \centering
  \includegraphics[width=0.85\textwidth]{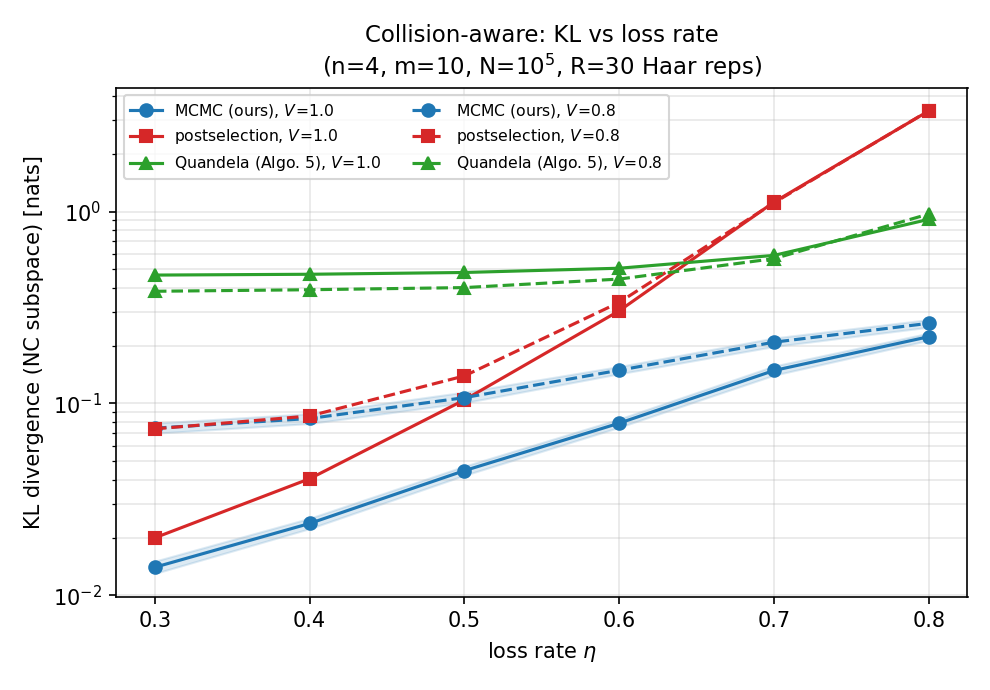}
  \caption{Collision-aware regime ($n=4, m=10$, $N=10^5$, $R=30$): KL on the NC subspace vs.\ $\eta$ under Perceval data. Solid $V=1.0$, dashed $V=0.8$. The Bayesian posterior (blue) is lowest at every $\eta$ except $\eta=0.3$, $V=0.8$, where postselection is marginally lower (Sec.~\ref{sec:4.3}); postselection (red) collapses at high loss; the recycling estimator (green) stays well above the posterior and cannot represent the collision components at all.}
  \label{fig:coll-kl-eta}
\end{figure}

Where Boson-Sampling output violates the no-collision restriction, the recycling family is structurally inapplicable: M\&M's NC-only kernel cannot consume samples that contain collisions. The collision-aware likelihood~\eqref{eq:coll-likelihood} handles them natively, and does so without the misspecification bias of the NC track: the jointly-inferred loss rate is \emph{exact} ($\hat\eta = 0.300/0.700/0.900$) at every $V$, because the collision-aware model is correctly specified for the collisions Perceval produces (Table~\ref{tab:coll-headline}). On the NC subspace (Fig.~\ref{fig:coll-kl-eta}) the Bayesian posterior is far below postselection at $\eta\geq0.7$, where postselection collapses ($\mathrm{KL}\!\gtrsim\!1.1$), and below the recycling estimator at every $\eta$, which stays at $0.38$--$0.92$ nats. At low loss it is comparable to postselection: at $\eta=0.3$ it is lower at $V=1.0$ ($0.014$ against $0.020$), tied at $V=0.90$ and $0.85$, and marginally higher at $V=0.8$ ($0.0744$ against $0.0738$; paired-bootstrap $95\%$ CI of the difference $[0.0001, 0.0011]$, lower on $10$ of $30$ circuits). As on the headline track, the postselection KL at high loss mostly measures the $1/N$ floor of the KL protocol: since $71\%$ of the lossless output mass lies on collision patterns at $m=10$ (mean over the $30$ circuits), only about $240$ collision-free full-$n$ samples survive at $\eta=0.7$ and about $3$ at $\eta=0.9$, spread over $K_{\mathrm{NC}}=210$ patterns, so many of its estimates are exactly zero (Sec.~\ref{sec:4.nsweep}). Coverage against $p^*$ drops to $0.67$ in the low-loss, low-$V$ corner ($\eta=0.3$, $V=0.8$) and is $0.92$--$0.96$ for $\eta\geq0.7$ and for $V=1.0$. Against $p_V$ the coverage in that corner is $0.95$, the $V=1$ value, so the drop reflects the distinguishability shift exceeding the narrow low-loss intervals rather than a failure of the fit (App.~\ref{app:target-shift}).

Table~\ref{tab:coll-headline} scores all methods on the NC subspace, because the baselines estimate only NC probabilities. The collision-aware posterior also estimates the collision patterns, which carry $71\%$ of the lossless output mass here. We refitted it on new Perceval samples for the same circuits and scored it on all $715$ patterns, on the collision patterns alone and on the NC patterns alone (Table~\ref{tab:coll-patterns}). At $V=1.0$ the collision patterns are recovered at least as accurately as the NC ones (KL $0.012$, $0.117$ and $0.269$ at $\eta=0.3$, $0.7$ and $0.9$, against $0.66$ for a uniform guess), the total collision mass is estimated to within $0.02$ of its true value $0.709$, and the coverage on the collision patterns is $0.95$--$0.96$. At $V=0.8$ the same target shift as on the NC subspace applies: the partially distinguishable distribution bunches less (collision mass $0.668$ for $p_V$), so the posterior places less mass on collisions ($0.67$--$0.71$) and its coverage of $p^*$ drops at low loss.

\begin{table}[t]
\centering
\caption{Collision-aware metrics at $n=4, m=10, N=10^5$, KL on the NC subspace. Convention as Table~\ref{tab:nc-headline}. The inferred $\hat\eta$ is exact at every cell: the collision-aware likelihood is correctly specified for the collisions Perceval generates. At $\eta=0.3$, $V=0.8$ postselection is marginally lower ($0.0738$ against $0.0744$), so no entry is bold. At $\eta\geq0.7$ the postselection KL is dominated by the $1/N$ floor applied to its zero estimates.}
\label{tab:coll-headline}
\small
\begin{tabular}{ccccccc}
\toprule
$\eta$ & $V$ & \textbf{MCMC KL} & post KL & Algo.\ 5 KL & $\hat\eta$ & cov \\
\midrule
0.3 & 1.0 & $\mathbf{0.014\,[0.013,0.015]}$ & $0.020$ & $0.467$ & $0.300$ & $0.94$ \\
0.3 & 0.8 & $0.074\,[0.070,0.079]$ & $0.074$ & $0.384$ & $0.300$ & $0.67$ \\
0.7 & 1.0 & $\mathbf{0.149\,[0.141,0.157]}$ & $1.120$ & $0.591$ & $0.700$ & $0.96$ \\
0.7 & 0.8 & $\mathbf{0.209\,[0.198,0.219]}$ & $1.111$ & $0.568$ & $0.700$ & $0.92$ \\
0.9 & 1.0 & $\mathbf{0.320\,[0.306,0.336]}$ & $3.236$ & $0.924$ & $0.900$ & $0.96$ \\
0.9 & 0.8 & $\mathbf{0.332\,[0.320,0.345]}$ & $1.679$ & $0.833$ & $0.900$ & $0.96$ \\
\bottomrule
\end{tabular}
\end{table}

\begin{table}[t]
\centering
\caption{Accuracy of the collision-aware posterior on the collision patterns ($n=4$, $m=10$, $N=10^5$, $R=30$, refitted on new samples). KL against $p^*$ on all $715$ patterns, on the $505$ collision patterns and on the $210$ NC patterns, each renormalised on its subset; a uniform guess gives KL $0.63$ on all patterns and $0.66$ on the collision patterns. Collision mass: estimated total probability of the collision patterns (true value $0.709$). Coverage of the $95\%$ credible intervals on the collision patterns and on all patterns. No baseline estimates collision patterns.}
\label{tab:coll-patterns}
\small
\begin{tabular}{cccccccc}
\toprule
$\eta$ & $V$ & KL all & KL coll. & KL NC & coll.\ mass & cov.\ coll. & cov.\ all \\
\midrule
0.3 & 1.0 & 0.012 & 0.012 & 0.013 & 0.709 & 0.95 & 0.95 \\
0.3 & 0.8 & 0.057 & 0.045 & 0.073 & 0.669 & 0.70 & 0.67 \\
0.7 & 1.0 & 0.125 & 0.117 & 0.146 & 0.715 & 0.96 & 0.96 \\
0.7 & 0.8 & 0.174 & 0.157 & 0.207 & 0.676 & 0.93 & 0.92 \\
0.9 & 1.0 & 0.285 & 0.269 & 0.322 & 0.728 & 0.96 & 0.96 \\
0.9 & 0.8 & 0.301 & 0.289 & 0.331 & 0.707 & 0.96 & 0.96 \\
\bottomrule
\end{tabular}
\end{table}

\subsection{Uncertainty quantification and retrospective coverage diagnostics}
\label{sec:4.4}

The joint posterior returns credible intervals on both the reconstructed distribution $p$ and the loss rate $\eta$; no recycling or postselection baseline supplies either. These intervals quantify posterior uncertainty conditional on the assumed likelihood. They do not, by themselves, diagnose model misspecification when the true distribution is unknown. In the present simulations, however, the known truth permits a retrospective coverage assessment: the empirical coverage of the $95\%$ intervals on $p(s)$ --- the fraction of the $K$ pattern probabilities whose true value falls inside its interval --- is $0.88$--$0.96$ on the no-collision headline and $0.67$--$0.96$ on the collision-aware track, near nominal everywhere except the low-loss, low-$V$ corner. In that corner the KL of the point estimate remains small (Table~\ref{tab:coll-headline}), yet coverage of $p^*$ falls to $0.67$. The intervals there cover the partially distinguishable distribution $p_V$ that the data determine (coverage $0.95$, App.~\ref{app:target-shift}) and miss $p^*$ wherever the distance between the two exceeds their width, which happens at low loss, where the intervals are narrowest. The undercoverage of $p^*$ is nevertheless real for a user: on hardware, where $V$ is unknown and not modelled, the intervals are calibrated for $p_V$ rather than $p^*$, and their coverage of $p^*$ falls as they narrow with lower loss or larger $N$ (on the no-collision track at $\eta=0.3$, $V=0.8$: $0.83$ at $N=2\times10^5$, against $0.94$ at $V=1.0$; Sec.~\ref{sec:4.nsweep}). At $\eta\geq0.7$ the coverage of $p_V$ is $0.98$--$0.99$, so the intervals are somewhat conservative at high loss. This retrospective check requires ground truth and therefore should not be interpreted as a deployable self-diagnostic on hardware. The loss rate is recovered exactly on the collision-aware track; on the no-collision track it carries the collision-induced bias of Sec.~\ref{sec:4.nsweep} ($\hat\eta=0.342$ at $\eta=0.3$ and $0.716$ at $\eta=0.7$; App.~\ref{app:coll-m20}).

Point-estimate accuracy in KL remains the primary metric by which we and the recycling literature judge mitigation; the coverage numbers are reported as empirical frequencies at the specific Haar unitaries, $\eta$, $V$, and $N$ of the grid, not as a formal simulation-based-calibration (Talts-style~\cite{Talts2018SBC}) guarantee. With that scope stated, the framework is the only compared method to emit credible intervals at all, extending the credible-region methodology of~\cite{ShangNgEtAl2013Credible} and the Bayesian-posterior-as-error-bar precedent of~\cite{BlumeKohout2010BME} to photon-loss mitigation.

\section{Discussion}
\label{sec:5}

\subsection{Sources of recycling reconstruction error}
\label{sec:5.1}

The bias of photon-recycling estimators is structural, not numerical, and it has \emph{two distinct sources} that the $n=4$ benchmarks separate. It is important not to conflate them: only the second involves collisions.

The first is a \emph{reconstruction} bias that is present already in the no-collision regime. Recycling decomposes each recycled probability into the ideal term plus an interference term and then replaces that interference term with its expectation value; Mills and Mezher bound the resulting error and note that it is present in all of their recycling estimators (\cite{MillsMezher2024}, \S4.3), while the exponential-extrapolation variant's decay function is, in their words, ``primarily heuristic.'' They characterise the consequence as a residual bias in the infinite-sample limit, $|p_{\mathrm{mit}}(s) - p_{\mathrm{id}}(s)| = \epsilon_{\mathrm{mitbias},s}$ (\cite{MillsMezher2024}, \S6), independently of collisions. Our $n=4, m=20$, $N=10^5$ comparison is consistent with that account and shows lower error for the likelihood-based posterior, which avoids the interference reweighting. It does not demonstrate that the posterior error continues to decrease to zero as $N\to\infty$.

A second, \emph{distinct} bias appears only in the collision regime. The recycling kernel
\[
\pi^{\mathrm{NC}}_y(p_{\mathrm{NC}}, \eta) = \eta^{k}(1-\eta)^{n-k} \sum_{s \in \mathcal{G}_{\mathrm{NC}}(y)} p_{\mathrm{NC}}(s)
\]
assumes the lossless state $s$ lives in $\{0,1\}^m$. The true data-generating distribution $p^\ast$ instead has support on all states with $\sum_i s_i = n$, including collision states such as $(2,0,0)$ at $n=2$, $m=3$. Photons in such states are lost independently per mode, so the correct per-mode law is the binomial form of~\eqref{eq:per-mode-binomial} from Sec.~\ref{sec:3.1}, which generalises the recycling kernel by a missing product of binomial coefficients $\prod_i \binom{s_i}{y_i}$. Equation~\eqref{eq:nc-misspec} formalises the resulting model mismatch: $\pi^{\mathrm{NC}} \neq \pi^{\mathrm{full}}$, so an estimator built from the NC kernel can retain misspecification error as the sample size grows whenever the omitted collision mass is nonzero. Our own no-collision-track inference exhibits this mismatch at $n=4, m=20$ --- visible as the residual $\hat\eta = 0.342$ in Sec.~\ref{sec:4.2} --- whereas the collision-aware likelihood yields an exact point estimate of $\eta$ in the tested $n=4, m=10$ benchmark. Establishing the population-limit bias of either estimator would require a separate asymptotic analysis.

\begin{equation}
\label{eq:nc-misspec}
\pi^{\mathrm{NC}}_y(p_{\mathrm{NC}}, \eta) \,=\, \eta^k (1-\eta)^{n-k}\!\!\!\!\sum_{s \in \mathcal{G}_{\mathrm{NC}}(y)} p_{\mathrm{NC}}(s)
\ \neq\
\pi^{\mathrm{full}}_y(p_{\mathrm{full}}, \eta).
\end{equation}

The two mechanisms concern the estimator--model pair rather than numerical optimisation alone. In the collision-aware experiments of Sec.~\ref{sec:4.3}, Algorithm~5 (Quandela \texttt{photon\_recycling}), which shares the NC kernel and the interference approximation with the rest of the recycling family, stays at $0.38$--$0.92$ nats across the tested loss rates. The collision-aware posterior, which fits the forward model directly and restores the $\prod_i \binom{s_i}{y_i}$ factor, tracks far below and matches the simulated loss rate in this benchmark. These results isolate plausible structural sources of the observed gap, but they do not determine the estimators' respective $N\to\infty$ limits.

\subsection{When MCMC dominates classical baselines}
\label{sec:5.2}

Three regimes make the Bayesian Multinomial estimator strictly preferable. First, at high loss ($\eta \geq 0.5$), postselection collapses because its effective sample count $N_{\mathrm{eff}} \sim N(1-\eta)^n$ shrinks rapidly --- a \emph{variance} floor, distinct from the recycling \emph{bias} floor: for the $n=4, m=20$ headline at $\eta=0.7, N=10^5$, postselection keeps about $0.4\%$ of the samples (roughly $430$ for $K=4845$ patterns), so most of its estimates are exactly zero. Its KL of $2.25$ against the posterior's $0.28$ is therefore largely the clip--floor penalty on those zeros rather than a like-for-like accuracy ratio, and it grows with $N$ for that reason (Sec.~\ref{sec:4.nsweep}). Second, in the collision-rich regime (lossless collision fraction $f_{\mathrm{coll}}=0.71$ at $m=10$), the recycling family (M\&M Algorithms~1--5, including Algorithm~5 shipped as Quandela \texttt{photon\_recycling}) structurally cannot consume the collision samples and discards them, whereas MCMC uses all $N$ events. Third, only the Bayesian framework supplies credible intervals on both $p$ and $\eta$, with retrospective empirical coverage $0.88$--$0.96$ at the tested resolutions (Sec.~\ref{sec:4.4}); because this assessment uses known simulation truth, it evaluates interval reliability but does not provide an automatic misspecification alarm for hardware data. None of the recycling baselines produces a posterior for $\eta$; the Bayesian framework's joint $(p, \eta)$ inference is therefore unique among compared methods, regardless of whether competing methods take $\eta$ as input, absorb it as a nuisance, or estimate it internally without exposing the estimate.

\subsection{Relation to prior work}
\label{sec:5.3}

Our comparison is a direct head-to-head against Mills and Mezher's shipped implementation (Quandela \texttt{photon\_recycling}) under data from an \emph{independent} simulator, not a re-derivation of their internal benchmarks. Under this independent-tool evaluation at the headline budget $N=10^5$, the Bayesian posterior sits below both recycling and postselection at every loss rate we test on the no-collision headline (on the collision-aware track, postselection matches it at low loss), and it additionally supplies collision support (Sec.~\ref{sec:3.1}), empirical credible-interval coverage on $p$ and $\eta$ (Sec.~\ref{sec:4.4}), and lower observed reconstruction error at $N=10^5$ (Sec.~\ref{sec:4.2}). The relationship to recycling is subsumptive rather than contradictory: the no-collision kernel is the boundary case of our likelihood~\eqref{eq:coll-likelihood}, and recycling can remain a sensible point estimator where its reconstruction error is small relative to postselection's variance and neither collisions nor uncertainty quantification are required. The ordering depends on the sample budget (Sec.~\ref{sec:4.nsweep}): at $N\leq10^4$ the two are comparable, with recycling slightly lower in some cells, and from $N\approx2\times10^4$ the posterior is lower and the gap widens up to $N=2\times10^5$. Whether it persists asymptotically requires a population-limit analysis and a sampler that scales beyond that budget.

\subsection{Limitations}
\label{sec:5.4}

Four limitations bound the present scope. First, the loss model is uniform per-photon; real platforms exhibit mode-dependent loss and dark counts. The likelihood generalises directly to $\eta \in [0,1]^m$ via the per-mode binomial in~\eqref{eq:per-mode-binomial}, and a principled extension is straightforward. Second, our collision-aware likelihood assumes photon-number-resolving (PNR) detectors to register multiply occupied modes, whereas Mills and Mezher work in the no-collision regime precisely so that standard threshold detectors suffice~\cite{MillsMezher2024}. With threshold detectors, collision events would be registered as apparent loss, which neither of our likelihoods models; handling them requires a coarsened likelihood with a distinct edge structure. Third, multiphoton emission and partial photon distinguishability are not modelled in the likelihood. The former degrades the estimator gracefully (App.~\ref{app:g2}); the latter moves its target from $p^*$ to $p_V$ (App.~\ref{app:target-shift}), so recovering $p^*$ rather than $p_V$ requires a likelihood that represents $V$, which is future work. Fourth, the collision-aware likelihood is benchmarked against the baselines at $m=10$ and compared with the no-collision fit at $m=20$ (App.~\ref{app:coll-m20}); exact NUTS handled $K_{\mathrm{coll}}=8855$ at $N=10^5$, but larger pattern spaces are left to approximate inference (Sec.~\ref{sec:6}). All experiments are simulation-based. Perceval provides an independent implementation cross-check at $V=1$ and $g^{(2)}=0$. On the collision-aware track this slice shares the uniform-loss statistical model assumed by the likelihood; the no-collision track is misspecified even there through the omitted collision patterns, and the $V<1$ and $g^{(2)}>0$ slices add physical effects that neither likelihood models; at $V<1$ these change the target rather than the fit. Hardware validation, with~\cite{Maring2024Ascella} as the natural target platform, is therefore the next priority. Computationally, exact NUTS becomes challenging beyond $K\approx10^4$ and, at $K=4845$, beyond $N\approx10^5$, and at $K=715$ beyond $N\approx2\times10^5$--$5\times10^5$ depending on the loss rate (Sec.~\ref{sec:4.nsweep}); variational and amortised families along the lines of~\cite{NikoloskaSimeone2024SBI} are the natural next step and are treated in Sec.~\ref{sec:6}.

\section{Conclusion and Future Work}
\label{sec:6}

We presented a Bayesian framework for photon-loss mitigation in linear optical quantum circuits that avoids the interference-reweighting approximations used by recycling-based estimators and extends naturally to the collision-rich output regime where Mills--Mezher style methods structurally cannot operate. Treating raw lossy detector counts as Multinomial observations under per-photon Bernoulli loss yields a posterior over the joint $(p, \eta)$, sampled via NUTS on a sparse incidence representation that kept inference tractable for pattern spaces of dimension up to $K=8855$ in our runs. Across the two scaled configurations $(n, m) \in \{(4,20)\ \text{no-collision}, (4,10)\ \text{collision-aware}\}$, with data generated by an independent simulator (Perceval) rather than the inference model, the posterior has the lowest KL divergence at every tested loss rate against both recycling and postselection on the no-collision headline and, on the collision-aware track, everywhere except at low loss under partial distinguishability, where postselection matches it; it exhibits empirical 95\% credible-interval coverage in $[0.88, 0.96]$ on the no-collision headline, and jointly estimates the loss rate (matching the simulated value in the tested collision-aware benchmark). It is moreover the only compared method that remains valid in the collision regime and that emits credible intervals at all, and it degrades gracefully under the multiphoton emission its likelihood does not model; under partial distinguishability it targets the partially distinguishable distribution while staying closest to the ideal one among the compared methods. These results establish a finite-sample advantage at the tested budgets, not asymptotic unbiasedness; $N$-sweeps up to $2\times10^5$ (no-collision) and $5\times10^5$ (collision-aware) show the advantage over recycling growing with $N$, but the large-sample limit still requires a population-limit analysis and a sampler that scales beyond the budget at which exact NUTS begins to fail.

\paragraph{Future directions.} The pattern-space dimension $K = \binom{m+n-1}{n}$ scales combinatorially, and the NUTS sampler we employ becomes expensive beyond $K \sim 10^4$ and fails at large sample counts ($N\gtrsim2\times10^5$ at $K=4845$, and by $N=10^6$ at $K=715$). The most immediate extension is therefore approximate Bayesian inference: variational families~\cite{NikoloskaSimeone2024SBI}, normalising-flow posteriors, or amortised inference networks could push the method into the regime of contemporary hardware experiments at $n \geq 6$. Mean-field variational inference is known to under-estimate posterior variance and would erode the empirical-coverage property reported here; richer families that capture posterior correlations are therefore the more promising route. A second extension drops the uniform-loss assumption: real photonic devices exhibit mode-dependent loss and small but non-zero distinguishability between photons. Mode-dependent loss can be absorbed into the likelihood by replacing $\eta \in (0,1)$ with a mode-dependent vector, at a modest increase in inference cost. Distinguishability is harder: it leaves the loss channel unchanged and moves the lossless distribution to $p_V$, which an unstructured $p$ absorbs without trace (App.~\ref{app:target-shift}). Returning the target from $p_V$ to $p^*$ therefore requires a model that links the two, for example by computing $p$ from the unitary with a partial-distinguishability parameter, which ties the inference to a characterised device. Hardware validation on Quandela's Ascella platform~\cite{Maring2024Ascella} or comparable systems is the natural empirical follow-up, and active experiment design --- choosing input states and measurement settings to maximise posterior information per shot --- is a third direction enabled by having a tractable posterior in hand.

\paragraph{Code and data availability.} The code, the result files behind every reported number, and the figure scripts are available from the authors on request.

\appendix

\section{Implementation: sparse-edge NUTS}
\label{app:implementation}

This appendix collects the implementation details that make exact NUTS sampling of the posterior~\eqref{eq:posterior} tractable at the configurations reported in Sec.~\ref{sec:4}. The components are individually standard --- NUTS~\cite{HoffmanGelman2014NUTS,Betancourt2017HMC} as implemented in NumPyro~\cite{Phan2019NumPyro} on JAX~\cite{Bradbury2018JAX}, stick-breaking reparameterisation of the probability simplex, coordinate-list (COO) sparse incidence storage, and the \texttt{segment\_sum} primitive available in modern automatic-differentiation frameworks. Their composition for the recycling-kernel likelihood is what reduces the inner-loop cost from the prohibitive dense $O(K \cdot |\mathcal{Y}|)$ gather to $O(E + K)$ on the configurations of interest.

\subsection{Sparse incidence representation and inner loop}
\label{app:implementation.sparse}

Posterior~\eqref{eq:posterior} has $K + 1$ unknowns: $K$ probabilities ($K = K_{\mathrm{NC}}$ or $K_{\mathrm{coll}}$ depending on regime) subject to the simplex constraint, plus $\eta \in (0, 1)$. We sample with NUTS after a stick-breaking reparameterisation of $p$ to an unconstrained $\mathbb{R}^{K-1}$ space. All reported results use a single chain with 300 warm-up and 500 post-warm-up draws, so $\hat R$ is not available for the grid runs themselves. We therefore diagnosed the sampler separately on four representative conditions --- loss $\eta=0.9$ with $V=1.0$ and $\eta=0.3$ with $V=0.8$, on each of the no-collision ($n=4,m=20$) and collision-aware ($n=4,m=10$) tracks --- with three Haar seeds each, $N=10^5$, and the flat prior $\alpha=1$. Each dataset was sampled with four independent chains at the same budget (300 warm-up, 500 draws per chain), and again with a longer reference run (four chains of 1000 warm-up and 1000 draws). In all $12$ production-budget runs there were no divergences~\cite{Betancourt2017HMC} and no draws at the maximum tree depth; the rank-normalised split-$\hat R$~\cite{Vehtari2021Rhat,GelmanRubin1992} was at most $1.007$ for $\eta$ and at most $1.015$ for every pattern probability, with at most $0.3\%$ of the $K$ components above $1.01$; bulk and tail effective sample sizes were at least $566$ and $397$ for every component. The posterior mean of $\eta$ agreed with its closed-form Beta posterior (the likelihood factorises in $\eta$ and $p$) to within $1.8$ Monte Carlo standard errors. Relative to the reference run on identical data, the production budget changed the KL divergence by at most $0.001$ nats ($0.3\%$), the $95\%$ coverage by at most $0.006$, and $\hat\eta$ by at most $4\times10^{-5}$; the reference runs satisfied $\hat R\leq1.01$ and effective sample size $\geq 400$ for all components. These checks cover the stated conditions only; we did not diagnose every cell of Sec.~\ref{sec:4}. In a stress test at the most concentrated prior of the sensitivity study below ($\alpha=0.25$, $N=10^3$, $\eta=0.3$, $V=1.0$; three Haar seeds), the production budget did \emph{not} reach per-component convergence: $2$--$4\%$ of the $p$ components had $\hat R>1.01$ (maximum $1.068$) and $5$--$11\%$ had tail effective sample size below $400$ (minimum $22$). There were still no divergences, and the KL divergence and coverage matched the reference run to within $3\times10^{-4}$ nats and $0.0012$; the reference run satisfied $\hat R\leq1.01$ for all but $0.02\%$ of the components. Component-wise intervals in the small-$N$, small-$\alpha$ cells of Fig.~\ref{fig:prior} should therefore be read with this caveat, and we did not diagnose the other cells of that grid.

The dominant cost inside the NUTS leapfrog integrator is the log-likelihood evaluation, which sums $\pi_y(p, \eta)^{c(y)}$ over all observed $y$. A dense $|\mathcal{Y}| \times K$ gather would be prohibitive at paper scale ($K = 4845$, $|\mathcal{Y}_{\mathrm{NC}}| = 6196$ at $n=4, m=20$). Instead, we precompute the sparse incidence between outcomes and inputs once per $(n, m)$ and store it as a triplet $(Y_{\mathrm{idx}}, P_{\mathrm{idx}}, W) \in \mathbb{Z}^E \times \mathbb{Z}^E \times \mathbb{R}^E$: each entry indexes one $(y, s)$ pair with $s \in \mathcal{G}(y)$, and $W$ holds the edge weight --- unity throughout the NC case, and the product of binomial coefficients $\prod_i \binom{s_i}{y_i}$ in the collision-aware case. The two regimes therefore differ in $W$ alone; the rest of the inner loop is shared. The inner loop, summarised in Algorithm~\ref{alg:sparse-nuts}, is a weighted gather $g = W \odot p[P_{\mathrm{idx}}]$ followed by a single \texttt{segment\_sum} over $Y_{\mathrm{idx}}$ to obtain $\sum_{s \in \mathcal{G}(y)} W \cdot p[s]$ for every $y$, then a closed-form combination with the $\eta^k(1-\eta)^{n-k}$ factor.

\begin{algorithm}
\caption{Sparse log-likelihood evaluation in NUTS inner loop}
\label{alg:sparse-nuts}
\begin{algorithmic}[1]
\Require Current $(p, \eta)$;\ flat edge arrays $(Y_{\mathrm{idx}}, P_{\mathrm{idx}}, W) \in \mathbb{Z}^E \times \mathbb{Z}^E \times \mathbb{R}^E$;
\Statex \hphantom{Require } $k$-of-$y$ vector $k_y \in \mathbb{Z}_{\geq 0}^{|\mathcal{Y}|}$;\ count vector $c \in \mathbb{Z}_{\geq 0}^{|\mathcal{Y}|}$.
\Ensure $\log \Pr(\mathbf{c} \mid p, \eta)$ (and JAX-traced gradient).
\State $\ell_\eta \gets \log \eta;\quad \ell_{1-\eta} \gets \log(1-\eta)$
\State $g \gets W \odot p[P_{\mathrm{idx}}]$ \Comment{$g \in \mathbb{R}^E$; element-wise weighted gather}
\State $\sigma_y \gets \mathrm{segment\_sum}(g,\ Y_{\mathrm{idx}},\ |\mathcal{Y}|)$ \Comment{$\sigma_y[y] = \sum_{s \in \mathcal{G}(y)} W \cdot p[s]$}
\State $\ell_y \gets k_y \cdot \ell_\eta + (n - k_y)\cdot \ell_{1-\eta} + \log \sigma_y$
\State \Return $\sum_y c(y) \cdot \ell_y$
\end{algorithmic}
\end{algorithm}

JAX traces the entire inner step into a single XLA kernel whose shape depends only on $(n, m)$. The trace is cached and reused across the $N$, $\eta$ grid; only the integer count vector $\mathbf{c}$ changes between runs. Empirically, this representation accelerates the inner loop by roughly two orders of magnitude over a dense JAX implementation, which is what brings exact MCMC at $K = 4845$ into the regime of routine practice.

\subsection{Complexity and scalability}
\label{app:implementation.complexity}

The pattern-space sizes are $K_{\mathrm{NC}} = \binom{m}{n}$ and $K_{\mathrm{coll}} = \binom{m+n-1}{n}$, and the edge count $E = |Y_{\mathrm{idx}}|$ is bounded by $K \cdot |\mathcal{Y}|$ but in practice substantially sparser. Each NUTS leapfrog step costs $O(E + K)$ on the sparse representation, dominated by the gather and segment-sum in Algorithm~\ref{alg:sparse-nuts}; tree-doubling and warm-up overheads are absorbed into the leapfrog count. Memory is linear in $E$, allowing the entire incidence structure to live in cache for the configurations of interest.

In practice we have run NUTS on CPU cores for every grid in Sec.~\ref{sec:4}, with the largest at $K_{\mathrm{NC}} = 4845$ ($n=4, m=20$, NC) and $K_{\mathrm{coll}} = 8855$ ($n=4, m=20$, collision-aware; App.~\ref{app:coll-m20}). These bounds match the regime targeted by Mills and Mezher's reference experiments~\cite{MillsMezher2024} and by current Quandela hardware~\cite{Maring2024Ascella}. Beyond $K \approx 10^4$, and at large $N$ (Sec.~\ref{sec:4.nsweep}), exact NUTS becomes impractical and approximate alternatives---mean-field or low-rank variational families, normalising-flow posteriors, and simulation-based inference, reviewed in tutorial form for parameterised-quantum-circuit simulators by Nikoloska and Simeone~\cite{NikoloskaSimeone2024SBI}---are the natural successors. Sec.~\ref{sec:6} returns to the scaling roadmap toward larger $n$.

\subsection{Prior sensitivity}
\label{app:prior}

\begin{figure}[t]
  \centering
  \includegraphics[width=0.95\textwidth]{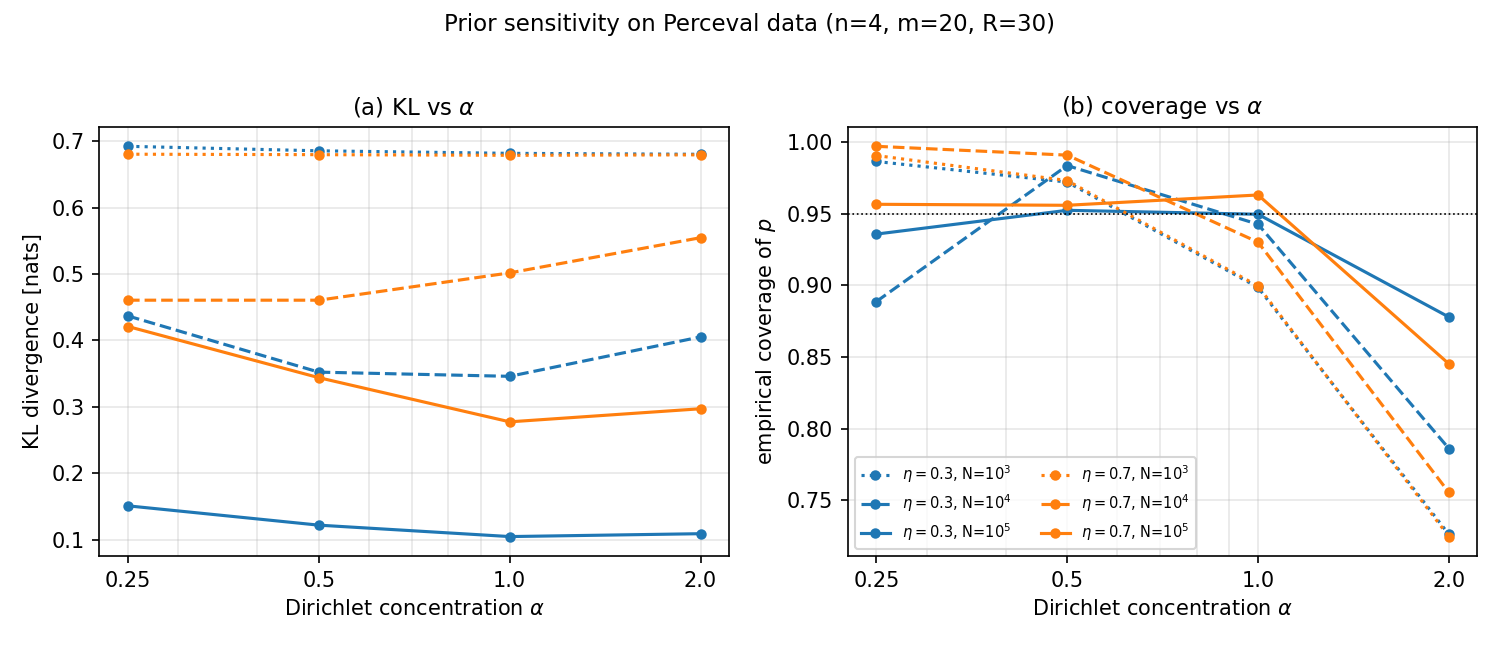}
  \caption{Prior sensitivity on Perceval-generated data at $n=4, m=20$ (NC), $R=30$ Haar replications, over the geometric grid $\alpha \in \{0.25, 0.5, 1, 2\}$ ($\log$ axis), by loss rate $\eta$ and sample size $N$. (a)~KL divergence is generally lowest near $\alpha = 1$ and rises at $\alpha = 2$ (most visibly at large $N$). (b)~$95\%$ credible-interval coverage varies non-monotonically for $\alpha \leq 1$, whereas $\alpha = 2$ under-covers in every tested cell; the dotted line marks the nominal $0.95$.}
  \label{fig:prior}
\end{figure}

The main results use a symmetric Dirichlet reference prior with concentration $\alpha = 1$ (uniform over the probability simplex). To assess sensitivity to this choice under the independent-simulator (Perceval) generator, we repeated the $n=4,\ m=20$ no-collision benchmark over a \emph{geometric} concentration grid $\alpha \in \{0.25, 0.5, 1, 2\}$ --- including the Jeffreys choice ($\alpha = 0.5$) and the flat choice ($\alpha = 1$), and spaced by factors of two because $\alpha$ is a multiplicative scale parameter --- at $\eta \in \{0.3, 0.7\}$ and $N \in \{10^3, 10^4, 10^5\}$, with $R=30$ replications and the NUTS settings of App.~\ref{app:implementation.sparse}. Figure~\ref{fig:prior} plots KL and credible-interval coverage against $\alpha$; Table~\ref{tab:prior-cov} reports the coverage surface.

Three points follow. First, the error metrics exhibit the expected shrinkage trade-off: larger $\alpha$ can reduce MSE by pulling estimates toward the simplex barycentre (shrinkage of the kind used in James--Stein-type estimators of multinomial frequencies~\cite{Hausser2009Shrinkage}), whereas KL --- our primary metric, following Mills and Mezher~\cite{MillsMezher2024} --- is generally lowest near $\alpha = 1$ and rises again at $\alpha = 2$, especially at high loss (Fig.~\ref{fig:prior}a). Second, coverage is not monotone in $\alpha$ in every cell; for example, at $\eta=0.3$ and $N=10^4$ it increases from $0.89$ at $\alpha=0.25$ to $0.98$ at $\alpha=0.5$. The consistent pattern is narrower: $\alpha=2$ under-covers in all tested cells ($0.72$--$0.88$), while $\alpha\leq1$ spans $0.89$--$1.00$ (Fig.~\ref{fig:prior}b, Table~\ref{tab:prior-cov}). Third, the inferred loss rate is essentially invariant to $\alpha$. We retain the flat choice $\alpha=1$ as a pragmatic balance between KL and empirical coverage on this grid, not as a universally optimal or uninformative prior. The $\alpha=0.5$ results provide a useful sensitivity comparison, while $\alpha=2$ illustrates the coverage cost of stronger shrinkage.

\begin{table}[t]
\centering
\caption{Empirical $95\%$ credible-interval coverage of $p$ by Dirichlet concentration $\alpha$ at $n=4, m=20$ (NC), Perceval data, $R=30$. Coverage is non-monotone across $\alpha \leq 1$; $\alpha = 2$ under-covers in every tested cell, while the flat choice $\alpha = 1$ yields coverage between $0.90$ and $0.96$.}
\label{tab:prior-cov}
\begin{tabular}{llcccc}
\toprule
$\eta$ & $N$ & $\alpha=0.25$ & $\alpha=0.5$ & $\alpha=1$ & $\alpha=2$ \\
\midrule
0.3 & $10^3$ & 0.99 & 0.97 & 0.90 & 0.73 \\
0.3 & $10^4$ & 0.89 & 0.98 & 0.94 & 0.79 \\
0.3 & $10^5$ & 0.94 & 0.95 & 0.95 & 0.88 \\
0.7 & $10^3$ & 0.99 & 0.97 & 0.90 & 0.72 \\
0.7 & $10^4$ & 1.00 & 0.99 & 0.93 & 0.76 \\
0.7 & $10^5$ & 0.96 & 0.96 & 0.96 & 0.85 \\
\bottomrule
\end{tabular}
\end{table}

\subsection{Multiphoton robustness}
\label{app:g2}

\begin{figure}[t]
  \centering
  \includegraphics[width=0.95\textwidth]{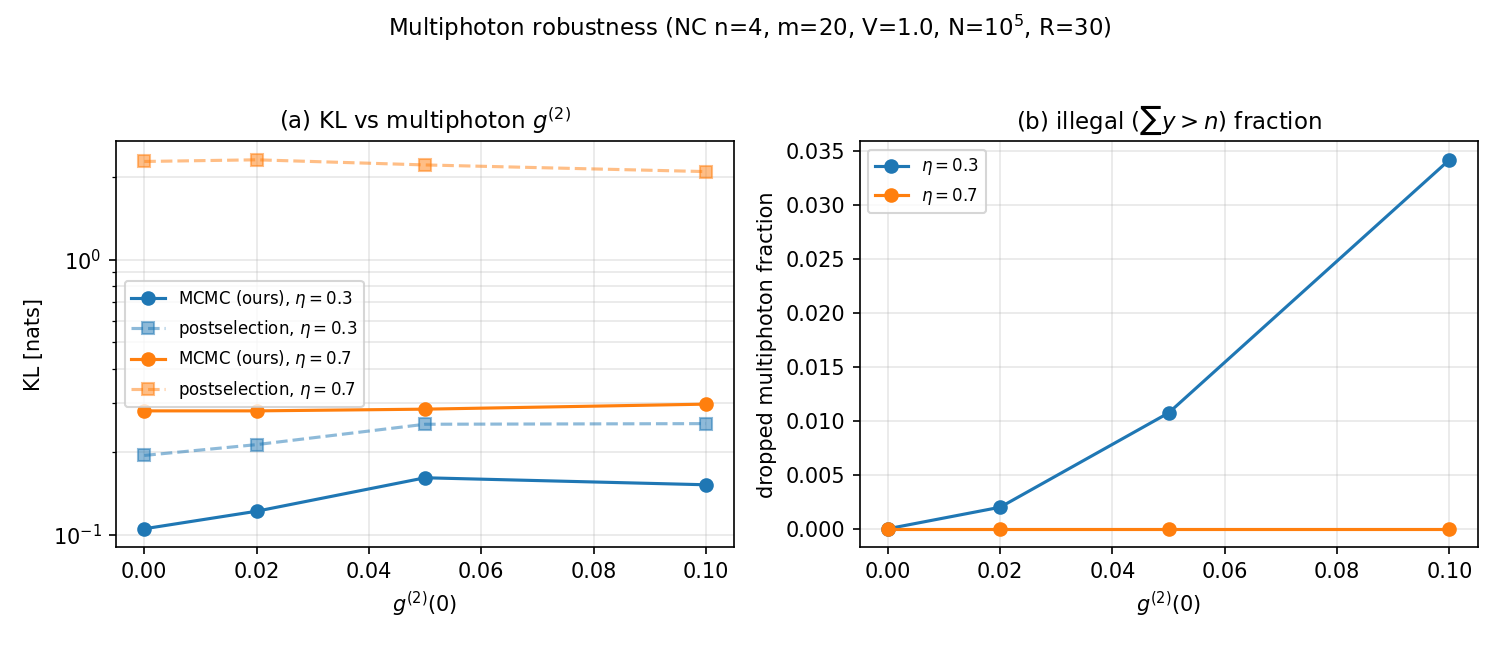}
  \caption{Multiphoton robustness at $n=4, m=20$ (NC), $V=1.0$, $N=10^5$, $R=30$. (a)~KL vs.\ $g^{(2)}$: the Bayesian posterior (solid) stays below postselection (dashed) at both loss rates; the effect of $g^{(2)}$ is confined to low loss ($\eta=0.3$). (b)~Fraction of $\sum_i y_i > n$ events discarded: it grows with $g^{(2)}$ at $\eta=0.3$ but stays at zero at $\eta=0.7$, where loss removes the surplus photons.}
  \label{fig:g2}
\end{figure}

The headline results fix $g^{(2)}=0$ (ideal single-photon sources). To probe robustness to multiphoton emission --- an effect the single-$\eta$ Binomial likelihood cannot represent, since it produces $\sum_i y_i > n$ events outside the model's support --- we swept $g^{(2)} \in \{0, 0.02, 0.05, 0.10\}$ at the $n=4, m=20$ no-collision headline ($V=1.0$, $N=10^5$, $R=30$) for $\eta \in \{0.3, 0.7\}$ (Fig.~\ref{fig:g2}). Two observations follow. First, $g^{(2)}$ degrades the estimator only at low loss: at $\eta=0.3$ the discarded multiphoton fraction rises to $3.4\%$ at $g^{(2)}=0.1$ and the posterior KL rises modestly ($0.105 \to 0.15$) with coverage dipping to $\sim0.89$, yet the posterior stays below both baselines throughout (recycling, not plotted, stays at $0.32$ at $\eta=0.3$ and $0.47$--$0.49$ at $\eta=0.7$). Second, at high loss ($\eta=0.7$) $g^{(2)}$ is essentially invisible --- the illegal fraction stays at zero and KL is flat --- because the surplus photon is itself usually lost, so loss suppresses multiphoton contamination. Even at $g^{(2)}=0.1$, well beyond typical single-photon-source values ($g^{(2)} \sim 0.01$--$0.03$), the estimator degrades gracefully rather than failing. Unlike partial distinguishability, $g^{(2)}>0$ is a genuine misspecification: it changes the number of photons rather than their distribution over modes, and $\hat\eta$ moves with it ($0.342\to0.321$ at $\eta=0.3$), whereas it is nearly unchanged by $V$ ($0.342\to0.338$ from $V=1$ to $0.8$; App.~\ref{app:target-shift}).

\section{Collision-aware likelihood at the headline size}
\label{app:coll-m20}

The no-collision headline fits the NC likelihood although, at $n=4$, $m=20$, bosonic bunching places a large share of the lossless output on collision patterns (about $45\%$ of the ideal output mass, averaged over the $30$ headline circuits). In the NC pipeline these events are either discarded, if a mode still holds two or more photons after loss, or, if loss has reduced them to a no-collision pattern, attributed to no-collision inputs. To measure what this costs, we fitted the collision-aware likelihood of Sec.~\ref{sec:3.1} ($K_{\mathrm{coll}}=\binom{23}{4}=8855$) on the same $30$ Haar unitaries and settings as the headline, with independent Perceval samples, and compared the two fits circuit by circuit (Table~\ref{tab:coll-m20}).

Three results follow. First, the collision-aware fit removes the loss-rate bias of the NC track ($\hat\eta=0.298$ and $0.697$ against $0.342$ and $0.716$ at $\eta=0.3$ and $0.7$, $V=1.0$), which identifies the collision mismatch as its source. Second, it does not lower the KL on the no-collision subspace: the KL is slightly higher, by $0.003$--$0.009$ nats, and the NC fit is lower on $16$--$25$ of the $30$ circuits. Third, it remains below both baselines on every circuit. The difference in KL is not a sampling artefact. Four-chain runs at the production budget on two circuits per loss rate ($V=1.0$) show no divergences and no draws at the maximum tree depth; $\hat R\leq1.011$ for $\eta$ and $\hat R\leq1.015$ for every pattern probability, with at most $0.2\%$ of components above $1.01$; tail effective sample size at least $343$ for every component; and KL values within $3\times10^{-4}$ nats of a four-chain reference run with 1000 warm-up and 1000 draws per chain.

The most likely explanation lies in three features of the model. The likelihood treats $p$ as an unstructured probability vector, so collision observations inform the collision-pattern probabilities but carry almost no information about the no-collision ones; the physical link between them, through permanents of a common unitary, is not used. A lossy outcome that looks collision-free can come from either kind of input: the collision-aware posterior splits it between them, whereas the NC fit assigns all of it to no-collision patterns, trading bias for concentration. Finally, the Dirichlet prior over $8855$ rather than $4845$ patterns shrinks more strongly. The collision-aware fit is therefore the appropriate model when the loss rate or the collision-pattern probabilities are of interest, quantities the no-collision estimators cannot provide, rather than a way to sharpen the estimate of the no-collision distribution.

\begin{table}[t]
\centering
\caption{Collision-aware vs.\ no-collision (NC) likelihood on the headline circuits ($n=4$, $m=20$, $N=10^5$, $R=30$ Haar unitaries, compared circuit by circuit). KL is on the NC subspace for both fits; $\Delta$KL $=$ collision-aware $-$ NC, with $95\%$ percentile-bootstrap CI. Coverage is over the $K_{\mathrm{coll}}=8855$ patterns for the collision-aware fit and over the $K_{\mathrm{NC}}=4845$ patterns for the NC fit. In every row the collision-aware posterior has lower KL than both recycling and postselection on all $30$ circuits.}
\label{tab:coll-m20}
\small
\setlength{\tabcolsep}{4pt}
\begin{tabular}{cccccccc}
\toprule
$\eta$ & $V$ & KL coll. & KL NC & $\Delta$KL & $\hat\eta$ coll. & $\hat\eta$ NC & cov.\ coll./NC \\
\midrule
0.3 & 1.0 & 0.108 & 0.105 & $+0.0028\,[+0.0011,+0.0044]$ & 0.298 & 0.342 & 0.951/0.949 \\
0.3 & 0.8 & 0.163 & 0.157 & $+0.0053\,[+0.0033,+0.0073]$ & 0.299 & 0.338 & 0.887/0.875 \\
0.7 & 1.0 & 0.289 & 0.280 & $+0.0090\,[-0.0017,+0.0198]$ & 0.697 & 0.716 & 0.959/0.962 \\
0.7 & 0.8 & 0.310 & 0.303 & $+0.0075\,[-0.0022,+0.0174]$ & 0.698 & 0.714 & 0.955/0.957 \\
\bottomrule
\end{tabular}
\end{table}

\section{Partial distinguishability as a shift of the target}
\label{app:target-shift}

Neither likelihood models the indistinguishability $V$, but the two effects that Perceval adds at $V<1$ and at $g^{(2)}>0$ differ in kind. Both likelihoods treat the lossless distribution $p$ as an unstructured probability vector over $n$-photon patterns and apply independent per-photon loss with a single $\eta$. Partial distinguishability changes where the $n$ photons go, through weaker interference, but not how many enter, and Perceval applies loss to each photon independently of its internal state. Data at $V<1$ are therefore uniform loss applied to the partially distinguishable lossless distribution $p_V$, which the likelihood represents exactly with $p=p_V$: partial distinguishability leaves the model correctly specified (apart from the omitted collisions of the no-collision track, which are present at every $V$), but makes its target $p_V$ rather than the ideal $p^*$. Multiphoton emission instead changes the number of photons, so that no $(p,\eta)$ reproduces the data, and $\hat\eta$ moves (App.~\ref{app:g2}). Two qualifications apply. The distinction relies on $p$ being unstructured; a likelihood that computed $p$ from permanents of the unitary could not represent $p_V$. It also relies on loss being independent of the photons' internal state, which holds in the simulation but may fail on hardware, for example for wavelength-dependent loss.

The reading makes three predictions, which we tested on the headline circuits. We computed $p_V$ exactly for each of the $30$ Haar unitaries with Perceval's SLOS backend at unit transmittance, without sampling (at $V=1$ it reproduces $p^*$ to within $10^{-8}$). First, the loss rate should be unaffected by $V$. It is: on the collision-aware track $\hat\eta$ is exact at every $V$ (Table~\ref{tab:coll-headline}), and on the no-collision track it barely moves ($0.342$ at $V=1$ and $0.338$ at $V=0.8$ for $\eta=0.3$), whereas $g^{(2)}=0.1$ moves it to $0.321$. Second, the increase of the posterior KL from $V=1$ to $V<1$ should match the distance $\mathrm{KL}(p^*\Vert p_V)$ between the two targets, which no estimator that ignores $V$ can remove. Table~\ref{tab:target-floor} compares the two on the no-collision subspace. On the collision-aware track at $\eta=0.3$ and $0.7$ they agree to within $0.004$ nats at every $V$. Where the estimation error is large ($\eta=0.9$, and $\eta=0.7$ on the no-collision track) the increase is smaller than the distance, since the two errors do not simply add; on the no-collision track at $\eta=0.3$ it is somewhat larger ($0.052$ against $0.044$ at $V=0.8$), possibly through an interaction with the omitted collisions, which we have not isolated.

\begin{table}[t]
\centering
\caption{Distance between the targets and increase of the posterior KL, on the NC subspace. $\mathrm{KL}(p^*\Vert p_V)$ is computed exactly (mean over the $30$ circuits); $\Delta\mathrm{KL}$ is the mean paired increase of the posterior KL against $p^*$ from $V=1$ to $V$ in the headline runs ($N=10^5$).}
\label{tab:target-floor}
\small
\begin{tabular}{lccccc}
\toprule
 & $V=1.0$ & $0.95$ & $0.90$ & $0.85$ & $0.80$ \\
\midrule
\multicolumn{6}{l}{\emph{Collision-aware track, $m=10$}} \\
$\mathrm{KL}(p^*\Vert p_V)$ & $0$ & $0.008$ & $0.023$ & $0.040$ & $0.060$ \\
$\Delta\mathrm{KL}$, $\eta=0.3$ & $0$ & $0.008$ & $0.023$ & $0.041$ & $0.060$ \\
$\Delta\mathrm{KL}$, $\eta=0.7$ & $0$ & $0.008$ & $0.026$ & $0.040$ & $0.060$ \\
$\Delta\mathrm{KL}$, $\eta=0.9$ & $0$ & $0.001$ & $0.003$ & $0.008$ & $0.012$ \\
\midrule
\multicolumn{6}{l}{\emph{No-collision track, $m=20$}} \\
$\mathrm{KL}(p^*\Vert p_V)$ & $0$ & $0.005$ & $0.015$ & $0.029$ & $0.044$ \\
$\Delta\mathrm{KL}$, $\eta=0.3$ & $0$ & $0.011$ & $0.023$ & $0.038$ & $0.052$ \\
$\Delta\mathrm{KL}$, $\eta=0.7$ & $0$ & $0.010$ & $0.015$ & $0.018$ & $0.023$ \\
$\Delta\mathrm{KL}$, $\eta=0.9$ & $0$ & $0.003$ & $0.003$ & $0.004$ & $0.004$ \\
\bottomrule
\end{tabular}
\end{table}

Third, the credible intervals should cover $p_V$, the distribution the data determine, at close to the $V=1$ rate. To test this we refitted both tracks at $V=0.8$ and at $V=1$ on new Perceval samples for the same circuits, with the production budget, and scored the same posterior against $p^*$ and against $p_V$ (Table~\ref{tab:target-cov}). Because the samples are new, the values against $p^*$ differ slightly from Tables~\ref{tab:nc-headline} and~\ref{tab:coll-headline}. Against $p_V$ the coverage is $0.95$--$0.99$ in every cell, including the cell where it is $0.67$ against $p^*$; the drop at low loss arises because the intervals there are narrower than the distance between $p_V$ and $p^*$ for about a fifth of the patterns. At $\eta\geq0.7$ the coverage against $p_V$ ($0.98$--$0.99$) is above nominal, so the intervals are somewhat conservative at high loss. The KL against $p_V$ is at or below its $V=1$ value; $p_V$ is flatter than $p^*$, since distinguishability weakens interference, and appears to be easier to estimate.

\begin{table}[t]
\centering
\caption{The same posterior scored against the ideal $p^*$ and against the partially distinguishable $p_V$ ($V=0.8$; $N=10^5$, $R=30$, refitted on new samples). Coverage is over the patterns of each fit; KL is on the NC subspace. The $V=1$ columns give the reference values, for which $p_V=p^*$.}
\label{tab:target-cov}
\small
\setlength{\tabcolsep}{4pt}
\begin{tabular}{llcccccc}
\toprule
 & & \multicolumn{3}{c}{Coverage} & \multicolumn{3}{c}{KL} \\
\cmidrule(lr){3-5}\cmidrule(lr){6-8}
Track & $\eta$ & vs.\ $p^*$ & vs.\ $p_V$ & $V=1$ & vs.\ $p^*$ & vs.\ $p_V$ & $V=1$ \\
\midrule
Collision-aware, $m=10$ & 0.3 & 0.67 & 0.95 & 0.95 & 0.074 & 0.012 & 0.014 \\
 & 0.7 & 0.92 & 0.98 & 0.96 & 0.207 & 0.105 & 0.146 \\
 & 0.9 & 0.96 & 0.99 & --- & 0.331 & 0.144 & --- \\
No-collision, $m=20$ & 0.3 & 0.87 & 0.97 & 0.95 & 0.157 & 0.090 & 0.105 \\
 & 0.7 & 0.96 & 0.99 & 0.96 & 0.308 & 0.175 & 0.283 \\
 & 0.9 & 0.95 & 0.99 & --- & 0.419 & 0.248 & --- \\
\bottomrule
\end{tabular}
\end{table}

\section*{Acknowledgements}

{\sloppy This work is in part supported by the National Research Foundation of Korea (NRF, \mbox{RS-2024-00451435}(20\%), \mbox{RS-2024-00413957}(20\%)), Institute of Information \& communications Technology Planning \& Evaluation (IITP, \mbox{RS-2025-02305453}(15\%), \mbox{RS-2025-02273157}(15\%), \mbox{RS-2025-25442149}(15\%) \mbox{RS-2021-II211343}(15\%)) grant funded by the Ministry of Science and ICT (MSIT), Institute of New Media and Communications(INMAC), and the BK21 FOUR program of the Education, Artificial Intelligence Graduate School Program (Seoul National University), and Research Program for Future ICT Pioneers, Seoul National University in 2026.\par}

\section*{Author contributions}

S.K.\ developed the methodology and software, performed the experiments and analysis, and drafted the manuscript. H.C., S.A.\ and D.N.\ contributed to the discussion of the results and revised the manuscript. J.L.\ supervised the project and revised the manuscript. All authors discussed the results and approved the final manuscript.

\emph{Use of AI.} The authors used Claude (Anthropic; Opus~4.7, Opus~5.5 and Sonnet~5.5) through the Claude Code tool for literature and citation checking, drafting and revising text, reformatting, writing and running the analysis code under the authors' supervision, and simulated pre-submission review. No AI was used to generate figures or data: every reported number comes from executed code. The authors reviewed all AI-assisted content and take full responsibility for the work.

\bibliographystyle{quantum}
\bibliography{references}

\end{document}